\documentclass[prd,preprint,superscriptaddress,preprintnumbers,eqsecnum,showpacs,nofootinbib,nobibnotes]{revtex4-1}
\usepackage{amsfonts,amsmath,amssymb,bm}
\allowdisplaybreaks[1] 
\usepackage[table]{xcolor}
\usepackage{graphicx} 
\usepackage{mathtools}
\usepackage{boondox-cal}
\usepackage{slashed}

\usepackage{xcolor}

\usepackage[utf8]{inputenc}

\newcommand{\be}{\begin{equation}}
\newcommand{\bea}{\begin{eqnarray}}
\newcommand{\ba}{\begin{align}}
\newcommand{\ee}{\end{equation}}
\newcommand{\eea}{\end{eqnarray}}
\newcommand{\ea}{\end{align}}

\definecolor{zero2}{rgb}{0.88,0.88,.88}

\def\1eq#1{Eq.~(\ref{#1})}

\def\2eqs#1#2{Eqs.~(\ref{#1}) and~(\ref{#2})}
\def\3eqs#1#2#3{Eqs.~(\ref{#1}),~(\ref{#2}) and~(\ref{#3})}
\def\4eqs#1#2#3#4{Eqs.~(\ref{#1}),~(\ref{#2}),~(\ref{#3}) and~(\ref{#4})}

\def\s#1{{\scriptscriptstyle #1}}

\def\G{\Gamma}

\def\s{\mathcal{s}}

\def\hphi0{{\hat\phi}_0}

\def\Mp2{{\mu}^2}

\def\extS0{\widetilde{\cal S}_0}

\makeatletter
\def\user@resume{resume}
\def\user@intermezzo{intermezzo}
\newcounter{previousequation}
\newcounter{lastsubequation}
\newcounter{savedparentequation}
\def\CT@@do@color{%
	\global\let\CT@do@color\relax
		\@tempdima\wd\z@
		\advance\@tempdima\@tempdimb
		\advance\@tempdima\@tempdimc
		\advance\@tempdimb\tabcolsep
		\advance\@tempdimc\tabcolsep
		\advance\@tempdima1.5\tabcolsep
	\kern-1.5\@tempdimb
	\leaders\vrule
	\hskip\@tempdima\@plus  1fill
	\kern-1.5\@tempdimc
	\hskip-\wd\z@ \@plus -1fill }
\makeatother

\usepackage[english]{babel}

\def\G{\Gamma}
\def\lh{{\ell_h}}
\def\lap{{\ell}_{a'}}

\begin{document}

\title{
Symmetry-Restoring Counter-terms\\
in Massive Gauge Theories with Non-Anticommuting $\gamma_5$\\
via Gauge-Invariant Sector Decomposition\\
}

\date{September 29th, 2026}

\author{A. Quadri}
\email{andrea.quadri@mi.infn.it}
\affiliation{INFN, Sezione di Milano, via Celoria 16, I-20133 Milano, Italy}

\begin{abstract}
\noindent
We compute the one-loop breaking of the Slavnov-Taylor (ST) identities 
in a massive Abelian chiral gauge theory with spontaneous symmetry breaking, regularized in the Breitenlohner-Maison-`t Hooft-Veltman (BMHV) scheme. We also provide the one-loop finite  counter-terms required to restore the one-loop ST identities.
The recent formulation of spontaneously broken models based on the Fr\"ohlich-Morchio-Strocchi (FMS) gauge-invariant and dynamical (i.e. propagating inside loops) fields is used.
The proposed algebraic approach holds true to all orders in the loop expansion and allows for a consistent decomposition of the ST breaking into independent sectors.
By using cohomological techniques we eventually constrain the dependence of the ST breaking on the vacuum expectation value of the Higgs field. This yields a set of differential equations that provide a useful check of the mass-dependent terms in the ST breaking terms in the broken phase.  They can be generalized to the non-Abelian case, both in the gauge-invariant formalism and in the conventional one.
In the massless limit we find agreement with previous calculations present in the literature.
\end{abstract}


\maketitle

\section{Introduction}

Renormalization of chiral gauge theories has recently regained some interest. The main reason is that, due to the presence of the $\gamma_5$ matrix, no invariant (i.e. respecting the relevant Slavnov-Taylor (ST) identities) regularization scheme is known.
Therefore one needs to recover the regularized broken ST identities, order by order in the loop expansion,
by adding suitably chosen finite counter-terms. 

The existence of such counter-terms in mathematically consistent regularization schemes like the Breitenlohner-Maison-`t Hooft-Veltman (BMHV) scheme was established long ago by using cohomological techniques and the locality properties of renormalizable theories, encoded in the so-called Quantum Action Principle (for a review see ~\cite{Belusca-Maito:2023wah,Piguet:1995er,Gomis:1994he}). 

The proof essentially makes use of the locality of the $n$-th order ST breaking term $\Delta^{(n)}$ if the ST identities have been restored up to order $n-1$ in the loop expansion and the fact that $\Delta^{(n)}$ obeys the so-called Wess-Zumino (WZ) consistency condition~\cite{Piguet:1995er,Gomis:1994he}.
This in turn implies that the problem of finding suitably chosen finite counter-terms $\Xi^{(n)}$ restoring the ST identity is equivalent to solving the equation
\begin{align}
    {\cal S}_0 (\Xi^{(n)}) = -\Delta^{(n)} \, ,
    \label{cts.cc}
\end{align}
subject to the WZ consistency condition 
\begin{align}
    {\cal S}_0 (\Delta^{(n)}) = 0 \, ,     
    \label{wz.cc}
\end{align}
${\cal S}_0$ being the linearized ST operator. Crucially ${\cal S}_0$ turns out to be nilpotent,
${\cal S}_0^2=0$
(a property that is in fact reflected into the WZ condition Eq.(\ref{wz.cc})) and therefore the classification of all possible solutions to Eq.(\ref{cts.cc})
amount to solving for the cohomology of ${\cal S}_0$ in the space of local action-like  functionals.
This problem has been solved in full generality and, if the gauge group is non-anomalous, the existence of $\Xi^{(n)}$ is guaranteed on purely algebraic grounds~\cite{Piguet:1995er,Gomis:1994he}.

On the other hand, the question of finding out the concrete expression of $\Delta^{(n)}$ for different physically relevant gauge groups is still partially an open problem.

Early attempts~\cite{Ferrari:1999nj} in the Algebraic Renormalization approach~\cite{Piguet:1995er} 
were based on the direct imposition of the ST identities. 
Namely, in spontaneously broken gauge theories the action-like part of the vertex functional (assuming that all particles are massive, so one can safely expand around zero momenta) is connected by the ST identities to the finite sector of the 1-PI amplitudes of dim. $\geq 5$. So the ST identities fix by themselves the symmetric action-like sector of the vertex functional, up to an ambiguity related to ${\cal S}_0$-invariants that are associated either to finite field redefinitions (for antifield-dependent invariants) or to the finite renormalization of the physical parameters of the theory (for cohomologically non-trivial invariants).
This in turn implies that the finite symmetry-restoring counter-terms can be worked out by direct comparison of the regularized action-like sector with its parameterization in terms of amplitudes of dim $\geq 5$ and of ${\cal S}_0$-invariant coefficients, to be fixed by imposing suitable normalization conditions.
Extension to the massless Yang-Mills theory in the BPHZL scheme was analyzed in~\cite{Quadri:2003pq} and to models with massive fermions in~\cite{Quadri:2005pv}.

Applications of the Algebraic Renormalization method in combination with the Background Field Method were proposed in~\cite{Grassi:2001zz}.
The one-loop renormalization of a general chiral gauge theory without scalar and Majorana fields was discussed in~\cite{Martin:1999cc}. The Abelian Higgs-Kibble model with chiral fermions was studied in~\cite{Sanchez-Ruiz:2002pcf}.

More recently, the full parameterization of the ST breaking insertion in the BMHV scheme was derived in a series of papers summarized in the review~\cite{Belusca-Maito:2023wah}. 
Standard Model (SM) symmetry-restoring counter-terms have been studied in
the symmetric unbroken phase ~\cite{Weisswange:2025tab}.
Higher-loop results were given in~\cite{Stockinger:2023ndm,Kuhler:2025znv,vonManteuffel:2025swv}.
Moreover, the general expression of the one-loop ST breaking in a massless chiral theory with fermions and scalars was presented in~\cite{Ebert:2024xpy}.
Non-abelian chiral gauge theories were considered in the background field method approach in~\cite{Cornella:2022hkc}.
In~\cite{OlgosoRuiz:2024dzq} an alternative approach was presented based on the use of the so-called spurion fields. Applications to the 
Standard Model Effective Theory (SMEFT) have been studied in~\cite{Fuentes-Martin:2025meq}.


In the present paper we apply the approach discussed in~\cite{Ebert:2024xpy} to an Abelian Higgs-Kibble model in the presence of $N_f$ families of chiral
fermions.
We retain the full mass dependence of the ST breakings and of the symmetry-restoring counter-terms, i.e. we derive the full dependence on the vacuum expectation value (v.e.v.) $v$ of the
physical Higgs field.

We work in the gauge-invariant formalism based on dynamical (i.e. propagating inside loops) Fr\"ohlich-Morchio-Strocchi fields~\cite{Quadri:2024aqo}.
The main feature of the latter formalism is that the ST identities decompose into separately invariant sectors, classified by the grading induced by the number of gauge-invariant fields of a given type circulating inside the loops. 

Thus it should be possible to obtain also independent ST breaking terms on a sector by sector basis. 
We will show that this in fact happens: both the ST breaking terms and finite counter-terms can be decomposed into independent sectors and organized according to the corresponding grading in the number of internal gauge-invariant fields.

We will compute explicitly at one loop order the relevant ST breakings in each sector. We work in the
BMHV scheme and amplitudes are evaluated in Landau gauge. The gauge choice is related to the particular simplicity of the results in Landau gauge, making the sector decomposition mechanism more transparent. The procedure can be extended in a straightforward way to other gauges.

We will also derive sector by sector the  counter-terms required to restore the ST identities 
at one loop order and then compare our results with those presented in~\cite{Ebert:2024xpy} for a generic Abelian massless model with scalars and fermions. The relevant ST breakings to be compared are given by the sum of all ST sectors contributions to a given symmetry-breaking insertion with fixed external legs. Our results coincide with those of Ref.~\cite{Ebert:2024xpy} in the massless limit $v \rightarrow 0$.

The sector decomposition of the ST breakings can be extended in a mathematically rigorous way recursively to all orders in the loop expansion. This might prove useful in inspecting specific sectors of the theory one is interested in, thus restricting the set of Feynman amplitudes that need to be evaluated while ensuring the consistent decoupling from different ST breaking sectors.

We also comment on the dependence of the ST breaking terms on the v.e.v.~$v$.
By using cohomological techniques based on the Algebraic Renormalization treatment of
the Background Field Method~\cite{Grassi:1995wr,Becchi:1999ir,Ferrari:2000yp}, we introduce an extended BRST differential transforming the v.e.v. into a constant anti-commuting ghost $\Omega$
\begin{align}
    s v = \Omega \, , \qquad s \Omega = 0 \, .
\end{align}
The extended WZ condition for the $n$-th loop ST breaking $\Delta^{(n)}$ allows one
to control the dependence of $\Delta^{(n)}$ on $v$ via a suitable differential equation
involving insertions of the physical Higgs scalar ans some suitable $\Omega$-dependent kernels.

In Landau gauge they become particularly simple  because the $\Omega$-dependent kernels vanish. In fact
one can reconstruct the full dependence of the ST breaking $\Delta^{(n)}$ on $v$ once
$\Delta^{(n)}$ is known in the massless limit $v \rightarrow 0$.
In more general cases they provide useful consistency checks of the computed ST breaking terms.

The paper is organized as follows. In Sect.~\ref{sec.classical action} we describe the classical action and the model field content. Moreover we describe the formulation of the theory based on the FMS fields. We comment on the definition of the vertex functional in $D$ dimensions and the simplest options to handle chiral fermions at the regularized level in the BMHV scheme.
In Sect.~\ref{sec.brst.sti} we describe how to paramterize the ST breaking insertion in the BMHV scheme for the concrete choice of the fermionic kinetic terms and fermion-gauge field interactions
presented in Sect.~\ref{sec.classical action}.
In Sect.~\ref{sec.1loop.st.brkgs} we provide the list of the explicit ST breaking terms at one loop order. They are classified according to the sector decomposition induced by the number of gauge-invariant fields.
In Sect.~\ref{sec.cts} we show how to derive the finite non-symmetric counter-terms once the ST breaking functional has been computed.
In Sect.~\ref{sec:comparison} we provide the comparison with existing results in the literature for massless chiral Abelian models. In Sect.~\ref{sec.wz} we prove that the dependence of the ST breaking on the v.e.v. $v$ is governed by the Wess-Zumino consistency condition of an extended BRST differential. Conclusions are presented in Sect.~\ref{sec.conclusions}.
Appendices are devoted to the BRST differentials of the theory, the functional identities of the model, the propagators and the explicit results of the ST breaking amplitudes, sector by sector in the gauge-invariant decomposition, as well as to the explicit parameterization of the symmetry-restoring counter-terms.

\section{Classical Vertex Functional in $D$ dimensions}\label{sec.classical action}

We consider an Abelian Higgs-Kibble model with $N_f$ chiral fermions and the usual quartic
Higgs potential. In four dimensions the classical action we start from is 
\begin{align}
S_{\mathrm{cl}}
 = \int d^4x\,\Big\{&
 -\frac{1}{4}F_{\mu\nu}F^{\mu\nu}
 +(D_\mu\phi)^\dagger D^\mu\phi
 -\frac{m^2}{2v^2}
  \left(\phi^\dagger\phi-\frac{v^2}{2}\right)^2
  +i\sum_{i=1}^{N_f}
  \bar\psi_i\gamma^\mu D_\mu\psi_i
 \nonumber\\[2mm]
&-\sum_{i,j=1}^{N_f}
 \left[
 K_{ij}\,\bar\psi_{L,i}\phi^\dagger\psi_{R,j}
 +K_{ij}^{*}\,\bar\psi_{R,j}\phi\psi_{L,i}
 \right]
\Big\}.
\label{cl.action}
\end{align}
In the above equation $F_{\mu\nu}$ is the field strength for the U(1) gauge field $A_\mu$, $F_{\mu\nu} = \partial_\mu A_\nu - \partial_\nu A_\mu$. The Higgs scalar field $\phi$ is
\begin{align}
    \phi=\frac{1}{\sqrt{2}}\left(v+\sigma+i\chi\right) \, ,
    \label{phi.sigma.chi}
\end{align}
$\sigma$ being the physical Higgs scalar and $\chi$ the pseudo-Goldstone field. $v$ is the Higgs vacuum expectation value (v.e.v.). We will sometimes also use the notation $\phi_0 = \sigma + v$.

The covariant derivatives are 
\begin{align}
D_\mu\phi
 =\left(\partial_\mu-ieY_S A_\mu\right)\phi\, , \qquad
D_\mu\psi_i
 =\left[
 \delta_{ij}\partial_\mu
 -ieA_\mu
 \left(
   (Y_L)_{ij}P_L+(Y_R)_{ij}P_R
 \right)
 \right]\psi_j,
 \label{cov.ders}
\end{align}
with chirality projectors
\begin{align}
P_L&=\frac{1-\gamma_5}{2},
\qquad
P_R=\frac{1+\gamma_5}{2}.
\label{chirality.projectors}
\end{align}
In the computations of the present paper we will consider diagonal matrices $Y_{L,R}$, $(Y_L)_{ij}=y_{L,i}\delta_{ij}, (Y_R)_{ij}=y_{R,i}\delta_{ij}$. We also choose the Yukawa coupling matrix $K_{ij}$ to be diagonal and real, 
$K_{ij} = K \delta_{ij}$.

The infinitesimal U(1) gauge transformations of parameter $\alpha$ are
\begin{equation}
\begin{gathered}
\delta_\alpha A_\mu
= \partial_\mu\alpha \, ,
\qquad
\delta_\alpha\psi_i
= ie\alpha
\left(
y_{L,i}P_L+y_{R,i}P_R
\right)\psi_i \, ,
\qquad
\delta_\alpha\overline{\psi}_i
= -ie\alpha\,\overline{\psi}_i
\left(
y_{L,i}P_R+y_{R,i}P_L
\right) \, ,
\\[2mm]
\delta_\alpha\phi
= ieY_S\alpha\,\phi \, ,
\qquad
\delta_\alpha\phi^\dagger
= -ieY_S\alpha\,\phi^\dagger \, .
\end{gathered}
\label{gt.1}
\end{equation}
or for the scalar components
\begin{align}
    \delta_\alpha\sigma =-eY_S\alpha\,\chi, \qquad
    \delta_\alpha\chi =eY_S\alpha\,(v+\sigma).
    \label{gt.2}
\end{align}
The scalar hypercharge $Y_S$ is related by gauge invariance to the fermionic hypercharges $Y_{L,R}$ by the condition
\begin{equation}
Y_LK-KY_R+Y_SK=0,
\end{equation}
that gives in components
\begin{align}
    Y_S = y_{R,i} - y_{L,i} \, .
    \label{yukawa.hypercharge}
\end{align}
In what follows we will set $Y_S=1$.
Moreover we also assume that the anomaly cancellation condition holds true:
\begin{equation}
\operatorname{Tr}\!\left(Y_R^3\right)
-\operatorname{Tr}\!\left(Y_L^3\right)
=
\sum_{i=1}^{N_f}
\left(y_{R,i}^3-y_{L,i}^3\right)
=0 \, .
\label{anomaly-cancellation}
\end{equation}

\subsection{Gauge-fixing}

The U(1) gauge symmetry in Eqs.(\ref{gt.1}) and (\ref{gt.2}) is promoted to the BRST symmetry 
of the gauge-fixed classical action by replacing the infinitesimal gauge parameter with the Abelian ghost $\omega$. We will work in the Landau gauge, so one 
adds to the classical action the gauge-fixing and ghost sectors
\begin{align}
S_{\mathrm{gf+gh}}
=\int d^4x\,\Big\{&
-b \partial A
+\bar\omega\,\Box\omega
\Big\} \, .
\label{landau.gf}
\end{align}
The BRST differential $s$ associated with the gauge group U(1) is reported in Eq.(\ref{U1.brst}). 
$S_{\mathrm{cl}}$~and $S_{\mathrm{gf+gh}}$ are separately invariant under $s$.

One also needs to introduce the sources coupled the non-linear BRST transformations of the quantum fields 
(the so-called anti-fields):
\begin{align}
S_{\mathrm{af}}
=\int d^4x\,\Big\{&
-e
\,\sigma^*\omega\chi
+e
\,\chi^*\omega(v+\sigma)
\nonumber\\
&+ie\sum_{i=1}^{N_f}
\bar\eta_i\,\omega
\left(
y_{L,i}P_L+y_{R,i}P_R
\right)\psi_i
-ie\sum_{i=1}^{N_f}
\omega\,\overline{\psi}_i
\left(
y_{L,i}P_R+y_{R,i}P_L
\right)\eta_i
\Big\}.
\label{antifield_sector}
\end{align}
As can be seen from Eq.(\ref{landau.gf}), in the Abelian case and in Landau gauge ghosts are free. This entails several simplifications since
the BRST transformations  do not get renormalized, 
i.e. the anti-field dependent amplitudes involving insertions of $\sigma^*,\chi^*,\eta_i, \bar\eta_i$ do not receive radiative corrections, since there are no Feynman diagrams contributing to those amplitudes. This will simplify considerably the subsequent analysis.
The full action $S_{\mathrm{cl}} + S_{\mathrm{gf+gh}} + S_{\mathrm{af}}$ is invariant under the gauge BRST differential $s$ in Eq.(\ref{U1.brst}).

\subsection{Gauge-invariant FMS fields}

We now turn to the introduction of the gauge-invariant FMS fields.
Let us define the composite gauge-invariant operators corresponding to the physical Higgs and the gauge fields~\cite{Quadri:2024aqo}:
\begin{align}
    \mathcal{H}(\phi)
\equiv
\frac{1}{v}
\left(
\phi^\dagger\phi-\frac{v^2}{2}
\right) \, , \qquad
\mathcal{A}_\mu(\phi,A)
\equiv
\frac{i}{ev^2}
\left[
2\phi^\dagger D_\mu\phi
-\partial_\mu\left(\phi^\dagger\phi\right)
\right].
\label{fms.fields}
\end{align} 
In the linearized approximation
\begin{align}
    \mathcal{H}(\phi) \sim \sigma + \dots \, , \qquad
    \mathcal{A}_\mu(\phi,A) \sim A_\mu - \frac{1}{ev} \partial_\mu \chi + \dots
\end{align}
so $\mathcal{H}(\phi)$ and $\mathcal{A}_\mu(\phi,A)$
are in 1-to-1 correspondence with the elementary fields $\sigma$
and the combination $ A_\mu - \frac{1}{ev} \partial_\mu \chi $.
Both of them are BRST-invariant under the linearized BRST operator, which is consistent with the fact that they represent physical gauge-invariant states.

The gauge-invariant formulation of spontaneously broken gauge theories proposed 
in~\cite{Quadri:2024aqo} 
promotes the above composite operators to independent dynamical fields, denoted by $h$ and $a_\mu$.
Their defining relations in Eq.(\ref{fms.fields}) are imposed off shell by Lagrange multipliers $X$ and $X_\mu$
respectively.
For that purpose one introduces the constraint sector
\begin{align}
S_{\mathrm{con}}
&=
\int d^4x\,\Bigg\{
X(\Box+m^2)
\left[
h-\frac{1}{v}
\left(
\phi^\dagger\phi-\frac{v^2}{2}
\right)
\right]
-\bar c(\Box+m^2)c
\nonumber\\[1mm]
&\hspace{2.8cm}
+X_\mu\Sigma^{\mu\nu}_{(0)}
\left[
a_\nu
-\frac{i}{ev^2}
\left(
2\phi^\dagger D_\nu\phi
-\partial_\nu(\phi^\dagger\phi)
\right)
\right]
-\bar c_\mu\Sigma^{\mu\nu}_{(0)}c_\nu
\Bigg\},
\label{S_constraint_full}
\end{align}
where the differential operator $\Xi^{\mu\nu}_{(0)}$ for the Landau gauge is 
\begin{equation}
\Xi^{\mu\nu}_{(0)}
=
\Box g^{\mu\nu}
-\partial^\mu\partial^\nu
+M_A^2 g^{\mu\nu} \, .
\end{equation}
The mass of the gauge field is given by $M_A = e v$.

Let us comment on Eq.(\ref{S_constraint_full}). By imposing the equations of motion for $X$
 and $X_\mu$
\begin{align}
& \frac{\delta S_{\mathrm{con}}}{\delta X} = (\Box+m^2)
\left[
h-\frac{1}{v}
\left(
\phi^\dagger\phi-\frac{v^2}{2}
\right)
\right] = 0 \, , \nonumber \\
& \frac{\delta S_{\mathrm{con}}}{\delta X_\mu} =
\Sigma^{\mu\nu}_{(0)}
\left[
a_\nu
-\frac{i}{ev^2}
\left(
2\phi^\dagger D_\nu\phi
-\partial_\nu(\phi^\dagger\phi)
\right)
\right] = 0
\end{align}
and neglecting zero modes of the operators $\Box+m^2$ and $\Sigma^{\mu\nu}_{(0)}$ (that do not contribute
perturbatively~\cite{Quadri:2024aqo}), one recovers the identification 
\begin{align}
h \sim \mathcal{H}(\phi) \, , \qquad a_\mu \sim \mathcal{A}_\mu(\phi,A) \, ,
\end{align}
i.e. the gauge-invariant fields $h, a_\mu$ do in fact coincide with the corresponding gauge-invariant
composite operators. 
Moreover one introduces scalar and vector ghost–antighost pairs $c,\bar c$ and 
$c_\mu, \bar c_\mu$, arranged into BRST doublets under an additional nilpotent constraint BRST differential $\s$:
\begin{align}
& \mathcal{s}\,\bar c
=
h-\frac{1}{v}
\left(
\phi^\dagger\phi-\frac{v^2}{2}
\right)\, ,
\qquad
\mathcal{s}\,X=c \, ,
\qquad
\mathcal{s}\,c=0 \, ,
\qquad
\mathcal{s}\,h=0 \, ,
\nonumber\\
& \mathcal{s}\,\bar c_\mu
=
a_\mu
-\frac{i}{e v^2}
\left[
2\phi^\dagger D_\mu\phi
-\partial_\mu\left(\phi^\dagger\phi\right)
\right] \, ,
\qquad
\mathcal{s}\,X_\mu=c_\mu \, ,
\qquad
\mathcal{s}\,c_\mu=0 \, ,
\qquad
\mathcal{s}\,a_\mu=0 \, . 
\end{align}
The constraint sector turns out to be $\s$-exact, namely
\begin{align}
    S_{\mathrm{con}} = \s \int d^4x \Big \{ \bar c (\Box + m^2) c + \bar c_\mu \Sigma^{\mu\nu}_{(0)} c_\nu \Big \}
\end{align}
and therefore cohomologically trivial. 
Thus the physical content of the theory remains unchanged.
Integrating out the multipliers and constraint ghosts via their equations of motion one
recovers the original formulation.
 The fermions remain elementary fields: only the scalar and gauge sectors are extended.

In order to define the non-linear constraint BRST transformations $\s \bar c, \s \bar c_\mu$,
the corresponding antifields $\bar c^*$ and $\bar c^*_\mu$ are required. Thus they are added to $S_{\mathrm{af}}$:
\begin{align}
S_{\mathrm{af}} \rightarrow
S_{\mathrm{af}} + \int d^4x \, 
\Bigg \{
\bar c^* \Big [ h-\frac{1}{v}
\left(
\phi^\dagger\phi-\frac{v^2}{2}
\right) \Big ] + 
\bar c^*_\mu \Big [
a_\mu
-\frac{i}{e v^2}
\left[
2\phi^\dagger D_\mu\phi
-\partial_\mu\left(\phi^\dagger\phi\right)
\right] 
\Big ] 
\Bigg \} \, .
\label{added.af}
\end{align}
As a final step, we notice that one can introduce a gauge-invariant mass operator for the field $h$
without spoiling power-counting renormalizability~\cite{Quadri:2024aqo}:
\begin{align}
    -\frac{M_h^2-m^2}{2}\,h^2
\end{align}
The dependence on the mass parameter $m$ cancels out from physical amplitudes~\cite{Quadri:2024aqo}, so that the physical mass parameter of the Higgs scalar is $M_h$. One can freely choose $m$ in order to simplify the computations. For that purpose in the present paper we choose
\begin{align}
m = M_h \, . 
\label{m.mH}
\end{align}
The four-dimensional vertex functional is finally given by (we denote by a bar purely four-dimensional quantities):
\begin{align}
    \overline{\G}^{(0)} = S_{\mathrm{cl}}
+
S_{\mathrm{gf+gh}}
+
S_{\mathrm{con}}
+
S_{\mathrm{af}}\, .
\end{align}
\subsection{$D$-dimensional definition of the vertex functional}

We are now in a position to describe the extension to $D$ dimensions in the BMHV regularization scheme.
In this scheme, the classical action is continued to
$D=4-2\epsilon$ dimensions by treating loop momenta and propagator
denominators as fully $D$-dimensional, while keeping $\gamma_5$
intrinsically four-dimensional. Every Lorentz covariant is decomposed
into a  four-dimensional part, denoted by a bar, and an
evanescent $(D-4)$-dimensional part, denoted by a hat. 
The $D=4$ and $D-4$-subspaces are mutually orthogonal, so that 
\begin{equation}
\gamma^\mu
=
\bar{\gamma}^{\mu}
+
\hat{\gamma}^{\mu},
\qquad
\bar g^{\mu\nu}\gamma_\nu
=
\bar{\gamma}^{\mu},
\qquad
\hat g^{\mu\nu}\gamma_\nu
=
\hat{\gamma}^{\mu},
\end{equation}
and
\begin{equation}
\bar{\gamma}^{\mu}\bar{\gamma}_{\mu}=4,
\qquad
\hat{\gamma}^{\mu}\hat{\gamma}_{\mu}
=D-4=-2\epsilon,
\qquad
\bar{\gamma}^{\mu}\hat{\gamma}_{\mu}=0.
\label{BMHV_decomposition}
\end{equation}
Chiral interaction vertices are consequently expressed in terms of
the $D=4$ covariants $\bar{\gamma}^{\mu}$, while possible terms
involving $\hat{\gamma}^{\mu}$ define evanescent extensions of the
four-dimensional action. Such terms vanish in the limit $D\to4$ but
can give finite contributions when multiplied by ultraviolet poles.
Cyclicity of the traces is preserved, while the $\gamma_5$ matrix obeys the following commutation and anticommutation relations:
\begin{equation}
\left\{\gamma_5,\bar{\gamma}^{\mu}\right\}=0,
\qquad
\left\{\gamma_5,\gamma^\mu\right\}
=
2\gamma_5\hat{\gamma}^{\mu},
\qquad
\left[\gamma_5,\hat{\gamma}^{\mu}\right]=0.
\label{BMHV_gamma5}
\end{equation}
Therefore, $\gamma_5$ anticommutes only with the 
four-dimensional Dirac matrices and commutes with their evanescent
counterparts.

When extending the classical action to $D$ dimensions, several options are possible~\cite{Ebert:2024xpy}. There is in fact no unique $D$-dimensional continuation of the fermionic sector. One
possibility is to combine the physical left- and right-handed fields into a single Dirac spinor.
Another possibility is to embed each physical chirality into a separate Dirac spinor, at the price of introducing sterile partners.

For practical computations we want to keep the number of fields to a minimum, so we use a single Dirac spinor. The drawback is that the evanescent part of the fermion kinetic term mixes chiralities carrying different hypercharges and can therefore break even global hypercharge invariance.

On top of that, the most general fermion-gauge interaction
contains evanescent couplings governed by the couplings $Y_{LR},Y_{RL}$~\cite{Ebert:2024xpy}:
\begin{align}
\mathcal{L}_{\mathrm{fermion\,int}} & =
-e Y_{R,i}\,
\overline{\psi}_i
P_L\overline{\slashed{A}}P_R\psi_i
-e Y_{L,i}\,
\overline{\psi}_i
P_R\overline{\slashed{A}}P_L\psi_i
\nonumber\\
&-e(Y_{RL})_{ij}\,
\overline{\psi}_i
P_L\widehat{\slashed{A}}P_L\psi_j
-e(Y_{LR})_{ij}\,
\overline{\psi}_i
P_R\widehat{\slashed{A}}P_R\psi_j .
\label{fermion_lagrangian_D}
\end{align}
In the above equation the gauge field $A_\mu$ is decomposed into its four dimensional
part $\overline{A}_\mu$ and its $D-4$-part
$\widehat{A}_\mu$. Hermiticity requires $Y_{RL}=Y_{LR}^{\dagger}$.

Terms in the second line of the above equation introduce an independent source of tree-level BRST breaking. At loop level, they produce additional evanescent Lorentz structures and substantially enlarge both the divergent and finite symmetry-restoring counter-terms. Some scalar–gauge and multi-gauge-boson counter-terms arise in fact entirely from these interactions~\cite{Ebert:2024xpy}.

In the present paper we therefore choose to set both $Y_{LR}$ and $Y_{RL}$ to zero (no tree-level interactions between fermions and the evanescent part of the gauge field) and to work with the standard kinetic fermion term in $D$
 dimensions
\begin{align}
\mathcal{L}_{\mathrm{fermion\,kin}} & =
i \bar \psi_i \slashed{\partial} \psi_i \, .
\label{fermion_lagrangian.kin.D} 
\end{align}
Since the fermionic kinetic term remain $D$-dimensional, 
chiralities with generally different hypercharges are mixed. This constitute the sole tree-level source of spurious BRST breaking induced by non-invariance of the BMHV regularization scheme.

We denote by $\G$ the $D$-dimensional symmetric vertex functional. $\G$ can be expanded according to the number of loops as
\begin{align}
    \G = \sum_{j=0}^{\infty} \G^{(j)} \, .
\end{align}
The classical D-dimensional vertex functional $\G^{(0)}$ obeys several functional identities that, unlike the  ST identities, are preserved
by the BMHV scheme,  since they are associated to symmetries that do not involve the fermionic fields.
We therefore assume that those identities are respected for $\G$ too.
The list of these identities is collected in Appendix~\ref{app.funct.ids}.

We would like to comment here on
Eqs.~\eqref{h_equation} and \eqref{amu_equation}.
In the symmetric (non-diagonal) basis
\[
\bigl(X,\sigma,h;\,A_\mu,\chi,b,X_\mu,a_\mu\bigr),
\]
the fields retain the form in which the functional identities are most
transparent, although the quadratic part of the action is not diagonal.
In particular, the gauge-invariant fields $h$ and $a_\mu$ do not possess
independent interaction vertices in this basis, as indicated by
Eqs.~\eqref{h_equation} and \eqref{amu_equation}. The mass-eigenstate
basis is obtained through linear field redefinitions that diagonalize
the quadratic action and make the physical propagating modes explicit;
in this basis, the gauge-invariant fields become interacting.

Instead of $a_\mu$ it is more convenient to use the gauge-invariant field 
$$ a'_\mu = a_\mu + \frac{1}{M_A^2} \partial_\mu b \, $$
since Feynman rules are simpler when expressed in terms of $a'_\mu$. So we will make use of $a'_\mu$ in our explicit computations.

 Consequently, 1-PI amplitudes can be graded by a pair $(\lh,\lap)$, where $\lh$ and $\lap$ denote, respectively, the numbers of internal $h$- and $a_\mu$-propagators.

The explicit propagators and the transformations relating the symmetric and mass-eigenstate bases are collected in Appendix~\ref{app.propagators}.

\section{ST identities and its breaking insertion}\label{sec.brst.sti}

The $D$-dimensional ST operator associated with the  BRST
differential $s$ is given by
\begin{align}
{\cal S}(\Gamma)
\equiv
\int d^Dx\,\Bigg\{&
\partial_\mu\omega\,
\frac{\delta\Gamma}{\delta A_\mu}
+
\frac{\delta\Gamma}{\delta\sigma^*}
\frac{\delta\Gamma}{\delta\sigma}
+
\frac{\delta\Gamma}{\delta\chi^*}
\frac{\delta\Gamma}{\delta\chi}
\nonumber\\
&+
\sum_{i=1}^{N_f}
\left[
\frac{\delta\Gamma}{\delta\bar\eta_i}
\frac{\delta\Gamma}{\delta\psi_i}
+
\frac{\delta\Gamma}{\delta\eta_i}
\frac{\delta\Gamma}{\delta\overline{\psi}_i}
\right]
+
b\,\frac{\delta\Gamma}{\delta\bar\omega}
\Bigg\} \, .
\label{U1_ST_identity}
\end{align}
As discussed above, $\G$ separately fulfills also at the regularized level the scalar and vector constraint ST identity in Eqs.(\ref{scalar_constraint_ST})
and (\ref{vector_constraint_ST}), thus we can safely focus on the gauge ST identity in Eq.(\ref{U1_ST_identity}).

Assuming that the ST identity has been fulfilled up to order $n-1$
in the loop expansion, the $n$-th order ST breaking reads
\begin{align}
\Delta^{(n)} = 
{\cal S}_0\!\left(\Gamma^{(n)}\right)
+\int d^Dx\,\Bigg\{&
\sum_{k=1}^{n-1}
\left[
\frac{\delta\Gamma^{(k)}}{\delta\sigma^*}
\frac{\delta\Gamma^{(n-k)}}{\delta\sigma}
+
\frac{\delta\Gamma^{(k)}}{\delta\chi^*}
\frac{\delta\Gamma^{(n-k)}}{\delta\chi}
\right]
\nonumber\\
&+
\sum_{i=1}^{N_f}\sum_{k=1}^{n-1}
\left[
\frac{\delta\Gamma^{(k)}}{\delta\bar\eta_i}
\frac{\delta\Gamma^{(n-k)}}{\delta\psi_i}
+
\frac{\delta\Gamma^{(k)}}{\delta\eta_i}
\frac{\delta\Gamma^{(n-k)}}
     {\delta\overline{\psi}_i}
\right]
\Bigg\},
\label{ST_order_n_linearized}
\end{align}
where  ${\cal S}_0$ is the linearized ST operator acting on a functional $F$ as
\begin{align}
{\cal S}_0(F)
\equiv
\int d^Dx\,\Bigg\{&
\partial_\mu\omega\,
\frac{\delta F}{\delta A_\mu}
+
b\,\frac{\delta F}{\delta\bar\omega}
\nonumber\\
&+
\frac{\delta\Gamma^{(0)}}{\delta\sigma^*}
\frac{\delta F}{\delta\sigma}
+
\frac{\delta F}{\delta\sigma^*}
\frac{\delta\Gamma^{(0)}}{\delta\sigma}
+
\frac{\delta\Gamma^{(0)}}{\delta\chi^*}
\frac{\delta F}{\delta\chi}
+
\frac{\delta F}{\delta\chi^*}
\frac{\delta\Gamma^{(0)}}{\delta\chi}
\nonumber\\
&+
\sum_{i=1}^{N_f}
\Bigg[
\frac{\delta\Gamma^{(0)}}{\delta\bar\eta_i}
\frac{\delta F}{\delta\psi_i}
+
\frac{\delta F}{\delta\bar\eta_i}
\frac{\delta\Gamma^{(0)}}{\delta\psi_i}
+
\frac{\delta\Gamma^{(0)}}{\delta\eta_i}
\frac{\delta F}{\delta\overline{\psi}_i}
+
\frac{\delta F}{\delta\eta_i}
\frac{\delta\Gamma^{(0)}}
     {\delta\overline{\psi}_i}
\Bigg]
\Bigg\}.
\label{linearized_ST_operator}
\end{align}
${\cal S}_0$ is nilpotent. 

Due to the Quantum Action Principle holding true in the BMHV scheme~\cite{Breitenlohner:1977hr}, $\Delta^{(n)}$ is known to be
a local functional in the fields, the external sources and their derivatives of dimension $\leq 5$ according to the canonical field dimensions given by Table~\ref{tab:uv_dimensions} in Appendix~\ref{app.propagators}.

It is also possible to determine explicitly the operator insertion giving rise to the ST breaking $\Delta^{(n)}$. 
For that purpose we follow the procedure of \cite{Ebert:2024xpy} and  compute the regularization-induced BRST breaking at tree level.
We start from the Lagrangian of the vertex functional $\overline{\G}^{(0)} = \int d^4x {\cal L}$, promote it to $D$-dimensions and evaluate its ST variation
as
\begin{align}
\widehat{\Delta}
&=
{\cal S}_D
\left(
\int d^Dx\,{\cal L}
\right)
=
-e\int d^Dx\,\omega\,
\overline{\psi}_i
\Bigg[
P_R
\left( y_{R,i} \,
\overleftarrow{\widehat{\slashed{\partial}}}
+
y_{L,i} \,
\overrightarrow{\widehat{\slashed{\partial}}}
\right)
P_R
\nonumber\\
&\hspace{35mm}
+
P_L
\left(
y_{R,i} \,
\overrightarrow{\widehat{\slashed{\partial}}}
+
y_{L,i} \,
\overleftarrow{\widehat{\slashed{\partial}}}
\right)
P_L
\Bigg]
\psi_i
\, .
\label{tree_level_BRST_breaking}    
\end{align}
Notice that the only source of the breaking in $\widehat{\Delta}$ are
the kinetic terms of the fermions, while there is no contribution from the Yukawa interactions. Thus $\widehat{\Delta}$ is not modified in the broken phase at  $v \neq 0$.

$\widehat{\Delta}$ controls the ST breaking terms, namely the functional $\Delta^{(n)}$ is generated by the insertion of the operator $\widehat \Delta$:
\begin{align}
\left . 
\frac{\delta^{(j)} \Delta^{(n)}}{\delta \Phi_1(p_1) \dots \delta \Phi_j(p_j) } \right |_{\Phi=0} =
\langle T \hat \Delta \cdot  \Phi_1(p_1) \dots \Phi_j(p_j)  \rangle^{\mathrm{1-PI}}
\end{align}
where $\Phi$ is a collective notation for the fields of the theory and in the r.h.s. of the above equation one only takes the 1-PI amplitudes.

We now notice that the 1-PI amplitudes in $\G^{(n)}$ can be graded according to the number of internal $h$ and $a'_\mu$-lines in the mass eigenstate basis. Let us denote by $\lh$ and $\lap$ those numbers. Then we can write
\begin{align}
    \G^{(n)}=\sum_{\lh,\lap}\G^{(n;\lh,\lap)} \, .
\end{align}
$\G^{(n;\lh,\lap)}$ is the generating functional of $n$-th order 1-PI amplitudes that contain in the loops $\lh$ internal propagators of the field $h$ and $\lap$ propagators of the field $a'_\mu$.
The breaking $\Delta^{(n)}$ can also be graded according to
the number of internal $\lh,\lap$ lines:
\begin{align}
\Delta^{(n;\lh,\lap)}
={}&
{\cal S}_0
\left(
\Gamma^{(n;\lh,\lap)}
\right)
\nonumber\\
&+
\int d^Dx\,
\sum_{k=1}^{n-1}
\sum_{{\cal j}_h=0}^{\lh}
\sum_{{\cal j}_{a'}=0}^{\lap}
\Bigg\{
\frac{\delta
\Gamma^{(k;{\cal j}_h,{\cal j}_{a'})}}
{\delta\sigma^*}
\frac{\delta
\Gamma^{(n-k;\lh-{\cal j}_h,\lap-{\cal j}_{a'})}}
{\delta\sigma}
\nonumber\\
&\hspace{28mm}
+
\frac{\delta
\Gamma^{(k;{\cal j}_h,{\cal j}_{a'})}}
{\delta\chi^*}
\frac{\delta
\Gamma^{(n-k;\lh-{\cal j}_h,\lap-{\cal j}_{a'})}}
{\delta\chi}
\nonumber\\
&\hspace{28mm}
+
\sum_{i=1}^{N_f}
\Bigg[
\frac{\delta
\Gamma^{(k;{\cal j}_h,{\cal j}_{a'})}}
{\delta\bar\eta_i}
\frac{\delta
\Gamma^{(n-k;\lh-{\cal j}_h,\lap-{\cal j}_{a'})}}
{\delta\psi_i}
\nonumber\\
&\hspace{46mm}
+
\frac{\delta
\Gamma^{(k;{\cal j}_h,{\cal j}_{a'})}}
{\delta\eta_i}
\frac{\delta
\Gamma^{(n-k;\lh-{\cal j}_h,\lap-{\cal j}_{a'})}}
{\delta\overline{\psi}_i}
\Bigg]
\Bigg\}.
\label{ST_order_n_lh_lap}
\end{align}
At one loop level only the ${\cal S}_0$-term in the r.h.s. of Eq.(\ref{ST_order_n_lh_lap}) is relevant.

\section{One-loop ST breaking in the BMHV scheme}\label{sec.1loop.st.brkgs}

We now set up our framework for the evaluation of 
$\Delta^{(1;\lh,\lap)}$.
Explicit computation have been carried out
by using FeynArts~\cite{Hahn:2000kx} for the generation of amplitudes and FeynCalc~\cite{Mertig:1990an,Shtabovenko:2023idz,Shtabovenko:2020gxv,Shtabovenko:2016sxi} for their evaluation. 

We can limit ourselves to amplitudes involving fermion loops or open fermionic chains, otherwise the insertion of $\widehat{\Delta}$ in Eq.(\ref{tree_level_BRST_breaking}) gives zero.

In the first class we find all amplitudes with the insertion of one ghost and up to four external scalar or vector legs. In the second class there are amplitudes with one ghost, a pair of fermion-antifermion fields and one scalar or one vector field.

It is convenient to use the symmetric basis parameterizing the external legs of the one-loop ST breaking. This is because, since $h$ and $a_\mu$ are gauge-invariant, there cannot be breakings with the insertion of one $h$ or $a_\mu$ external leg, as a consequence of Eqs.(\ref{g.inv.indep.symm.basis}). Internal loops are instad expressed in terms of the diagonal mass eigenstates.

Furthermore, by virtue of the $X$ and $X_\mu$ equations (\ref{X_equation}) and (\ref{Xmu_equation}), amplitudes involving at least one $X$ or one $X_\mu$ legs are not independent. They instead can be derived algebraically by taking the appropriate functional derivatives of 
Eqs.(\ref{X_equation}) and (\ref{Xmu_equation}), by which one insertion of $X$ is traded for the insertion of the external source $\bar c^*$ and
one insertion of $X_\mu$ is replaced by the insertion of the
external source $\bar c^{*\mu}$. As can be seen from Eq.(\ref{added.af}), there are no Feynman rules involving those sources and the fermions and thus at one loop level there are no Feynman diagrams with a fermionic loop contributing to the insertion of $\widehat \Delta$ and one or more $\bar c^*,\bar c^{*\mu}$ sources.
Thus one only needs to compute the one loop ST breaking with the insertion of a ghost, a pair of fermion and antifermion fields and a source $\bar c^*$ or $\bar c^{*\mu}$.

We start listing amplitudes with the insertion of a ghost and one or more scalars ore vectors. The study of amplitudes with fermionic external legs 
will be presented in Sect.~\ref{sec.delta.fermions}.

We denote the finite part of the amplitudes by the subscript $0$, while the coefficient of $\frac{1}{D-4}$  in the Laurent expansion around $D=4$ is denoted by the subscript $1$. The superscripts $\lh,\lap$ denote the sector to which the breaking belongs. For instance
$$ \Delta^{(1;0,0)}_{0;\omega(-p)\chi(p)}$$
is the one-loop finite part of the breaking with one ghost $\omega$ and one Goldstone $\chi$ field in the sector $(0,0)$, i.e. the sum of amplitudes with an external ghost and an external Goldstone leg with no $h$- and $a'_\mu$-internal propagators. The sum of all contributions from all sector is denoted by omitting the $\lh,\lap$ superscripts, e.g. $\Delta^{(1)}_{0;\omega(-p)\chi(p)}$ is the sum of all amplitudes with one external ghost and one external Goldstone $\chi$ legs. 

In the breaking sectors involving a pure ghost, a ghost with scalar legs, a ghost with vector legs, or a ghost with both scalar and vector legs, the only contributions arise from closed internal fermion loops belonging to the invariant sector $(0,0)$. We therefore omit the corresponding superscript indices from these amplitudes.

If not otherwise indicated, $q$ denotes the momentum of the ghost. By power-counting we only need to consider amplitudes with dimension $\leq 5$.

\subsection{One ghost}

The breaking $\Delta^{(1)}_{\omega(0)}$ vanishes.

\subsection{One ghost and one scalar}

\begin{equation}
\Delta^{(1)}_{0;\,\omega\!\left(-p\right)\,\chi(p)}
=
\frac{e K m_{e}}{12\pi^{2}}
\left(\mathrm{Tr} (Y_L) -\mathrm{Tr} (Y_R)\right)
\left(\bar{p}^{\,2}-6m_{e}^{2}\right)\, , 
\label{c_chi.0}
\end{equation}
\begin{equation}
\Delta^{(1)}_{1;\,\omega\!\left(-p\right)\,\chi(p)}
=
\frac{eK\,m_e\,\hat{p}^{\,2}}{12\pi^2}
\left(\mathrm{Tr} (Y_R)-\mathrm{Tr} (Y_L)\right).
\label{c_chi.1}
\end{equation}
$\Delta^{(1)}_{0;\,\omega\!\left(-p\right)\,\sigma(p)}$ and 
$\Delta^{(1)}_{1;\,\omega\!\left(-p\right)\,\sigma(p)}$ both vanish.

\subsection{One ghost and two scalars}

In this sector $\Delta^{(1)}_{0;\omega(q)\chi(p_1)\chi(p_2)}$, $\Delta^{(1)}_{1;\omega(q)\chi(p_1)\chi(p_2)}$
both vanish, as well as
$\Delta^{(1)}_{0;\omega(q)\sigma(p_1)\sigma(p_2)}$, $\Delta^{(1)}_{1;\omega(q)\sigma(p_1)\sigma(p_2)}$.
The mixed ghost-scalar-Goldstone breakings instead are
\begin{equation}
\Delta^{(1)}_{0;\,
\omega\!\left(q\right)\,
\chi(p_{1})\,\sigma(p_{2})}
=
\frac{eK^2}{12\pi^2}
\left( \mathrm{Tr} (Y_R)-\mathrm{Tr} (Y_L) \right)
\left(
18m_e^2-\bar p_1^{\,2}-3\bar p_2^{\,2}
-3\bar p_1\cdot\bar p_2
\right) \, ,
\label{c_chi_sigma.0}
\end{equation}
\begin{equation}
\Delta^{(1)}_{1;\,
\omega\!\left(q\right)\,
\chi(p_{1})\,\sigma(p_{2})}
=
\frac{eK^2}{12\pi^2}
\left(\mathrm{Tr} (Y_R) -\mathrm{Tr} (Y_L) \right)
\left(\hat p_1^{\,2}-3\hat p_2^{\,2}\right).
\label{c_chi_sigma.1}
\end{equation}
\subsection{One ghost and three scalars}

The non-vanishing amplitudes in this sector are
\begin{equation}
\Delta^{(1)}_{0;\,
\omega\!\left(q\right)\,
\chi(p_{1})\,\chi(p_{2})\,\chi(p_{3})}
=
\frac{eK^3m_e}{\pi^2}
\left(\mathrm{Tr} (Y_R) -\mathrm{Tr} (Y_L) \right)\, ,
\label{c_chi_chi_chi.0}
\end{equation}
\begin{equation}
\Delta^{(1)}_{0;\,
\omega\!\left(q\right)\,
\chi(p_{1})\,\sigma(p_{2})\,\sigma(p_{3})}
=
\frac{3eK^3m_e}{\pi^2}
\left(\mathrm{Tr} (Y_R) -\mathrm{Tr} (Y_L)\right).
\label{c_chi_sigma_sigma.0}
\end{equation}

\subsection{One ghost and four scalars}

The non-vanishing amplitudes in this sector are
\begin{equation}
\Delta^{(1)}_{0;\,
\omega\!\left(q\right)\,
\chi(p_{1})\,\chi(p_{2})\,\chi(p_{3})\,\sigma(p_{4})}
=
\frac{eK^4}{\pi^2}
\left(\mathrm{Tr} (Y_R) -\mathrm{Tr} (Y_L) \right) \, ,
\label{c_chi_chi_chi_sigma.0}
\end{equation}
\begin{equation}
\Delta^{(1)}_{0;\,
\omega\!\left(q \right)\,
\chi(p_{1})\,\sigma(p_{2})\,\sigma(p_{3})\,\sigma(p_{4})}
=
\frac{3eK^4}{\pi^2}
\left(\mathrm{Tr} (Y_R) -\mathrm{Tr} (Y_L) \right).
\label{c_chi_sigma_sigma_sigma.0}
\end{equation}

\subsection{One ghost and mixed scalar-vector fields}

There are no non-vanishing contributions to the  ST breaking with one ghost, one scalar and the vector field.
In the sector with one vector and two scalars the non-vanishing contributions are
\begin{equation}
\Delta^{(1)}_{0;\,\omega\!\left(q\right)\,
\chi(p_{1})\,\chi(p_{2})\,A_{\mu(p_{3})}}
=
\frac{i e^2K^2}{4\pi^2}
\mathrm{Tr} (Y_L - Y_R)^2
\left(
2\bar p_1^{\mu}
+2\bar p_2^{\mu}
+\bar p_3^{\mu}
\right) \, ,
\label{c_chi_chi_A.0}
\end{equation}
\begin{equation}
\Delta^{(1)}_{0;\,
\omega\!\left(q\right)\,
\sigma(p_{1})\,\sigma(p_{2})\,A_{\mu}(p_{3})}
=
\frac{i e^2K^2}{4\pi^2}
\mathrm{Tr} (Y_L - Y_R)^2
\bar p_3^{\mu}.
\label{c_sigma_sigma_A.0}
\end{equation}
There are no non-vanishing amplitudes with one ghost, one vector and three scalars.

There is only one non-vanishing amplitude with the insertion of a ghost, a scalar and two vectors, namely
\begin{equation}
\Delta^{(1)}_{0;\,
\omega\!(q)\,
\chi(p_{3})\,A_{\mu_{2}}(p_{2})\,A_{\mu_{1}}(p_{1})}
=
\frac{e^3K m_e}{2\pi^2}
\mathrm{Tr} (Y_L - Y_R)^3
g^{\mu_1\mu_2}.
\label{c_chi_A_A.0}
\end{equation}
There is also only one non-vanishing amplitude with one ghost,two scalars and two vectors:
\begin{equation}
\Delta^{(1)}_{0;\,
\omega\!\left(q\right)\,
\chi(p_{4})\,\sigma(p_{3})\,
A_{\mu_{1}}(p_{1})\,A_{\mu_{2}}(p_{2})}
=
\frac{e^3K^2}{2\pi^2}
\mathrm{Tr} (Y_L - Y_R)^3
g^{\mu_1\mu_2}.
\label{c_chi_sigma_A_A.0}
\end{equation}
The ST breakings with one ghost, one scalar and three vector legs are zero.

\subsection{One ghost and vector fields}

We report the non-vanishing amplitudes with one ghost and vector external legs as follows:
\begin{itemize}
    \item one vector:
    \begin{equation}
    \Delta^{(1)}_{0;\,
    \omega\!\left(-p\right)\,A_{\mu}(p)}
    =
    \frac{i e^2}{48\pi^2}
    \mathrm{Tr} (Y_L - Y_R)^2
    \left(6m_e^2-\bar p^{\,2}\right)
    \bar p^{\mu} \, ,
    \label{c_A.0}
    \end{equation}
\begin{equation}
     \Delta^{(1)}_{1;\,
    \omega\!\left(-p\right)\,A_{\mu}(p)}
    =
    -\frac{i e^2}{24\pi^2}
    \mathrm{Tr} (Y_L + Y_R)^2
    \hat p^{\,2}\,\bar p^{\mu}.
    \label{c_A.1}
\end{equation}
\item two vectors:
\begin{equation}
\Delta^{(1)}_{0;\,
\omega\!(q)\,
A_{\mu_{1}}(p_{1})\,A_{\mu_{2}}(p_{2})}
=
\frac{e^3}{12\pi^2}
\Big ( 
\mathrm{Tr} (Y_L)^3 - \mathrm{Tr} (Y_R)^3
\Big )
\varepsilon^{\mu_1\mu_2\rho\sigma}
\bar p_{1\rho}\bar p_{2\sigma}.
\label{c_A_A.0}
\end{equation}
The r.h.s. of the above equation is zero due to the
anomaly cancellation condition on the fermion content of the model in Eq.(\ref{anomaly-cancellation}).
\item three vectors:
\begin{equation}
\begin{aligned}
&\Delta^{(1)}_{0;\,
\omega\!\left(q\right)\,
A_{\mu_{1}}(p_{1})\,A_{\mu_{2}}(p_{2})\,A_{\mu_{3}}(p_{3})}
\\[2mm]
&\quad=
-\frac{i e^4}{24\pi^2}
\mathrm{Tr}(Y_L - Y_R)^4
\Big[
g^{\mu_{2}\mu_{3}}
\left(\bar p_{1}+\bar p_{2}+\bar p_{3}\right)^{\mu_{1}}
\\
&\qquad\qquad
+g^{\mu_{1}\mu_{3}}
\left(\bar p_{1}+\bar p_{2}+\bar p_{3}\right)^{\mu_{2}}
+g^{\mu_{1}\mu_{2}}
\left(\bar p_{1}+\bar p_{2}+\bar p_{3}\right)^{\mu_{3}}
\Big].
\end{aligned}
\end{equation}
\end{itemize}

\subsection{The fermionic sector}\label{sec.delta.fermions}

\begin{figure}[htbp]
    \centering
    \includegraphics[width=0.75\textwidth]
        {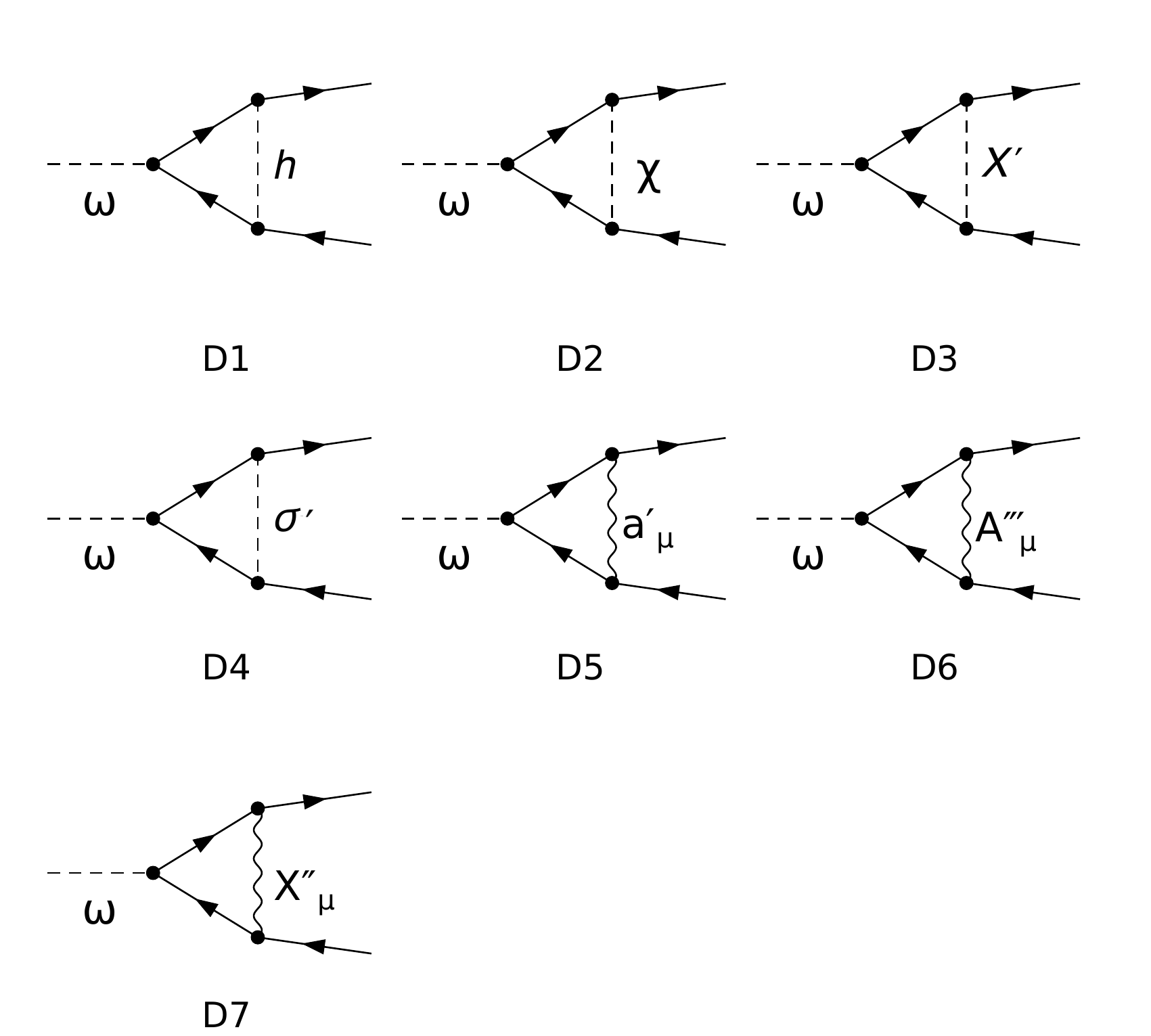}
    \caption{One-loop diagrams contributing to the ST breaking $\Delta^{(1)}_{\omega \psi_i \bar \psi_i}$.}
    \label{fig:c_f_f}
\end{figure}

The ST breaking with external fermion lines reveal a non-trivial sector decomposition.
Let us illustrate the main features on the case of the insertion of one ghost and the fermion-antifermion fields.
For a given species $\psi_i$ the relevant diagrams are collected in Figure~\ref{fig:c_f_f}.
The vertex to which the ghost $\omega$ is attached
is the ST breaking insertion in Eq.(\ref{tree_level_BRST_breaking}).

The interaction vertex fermion-antifermion-gauge field involves only the combination $A_\mu$, that is given in terms of the mass eigenstates used in the loop by Eq.(\ref{A_mass_eigenstate}):
\begin{align}
    A_\mu = A''_\mu + X''_\mu + a'_\mu - \frac{1}{M_A^2} \partial_\mu b'  \, .
\end{align}
It is convenient to introduce the combination
\begin{align}
    A'''_\mu = A''_\mu - \frac{1}{M_A^2} \partial_\mu b' \, ,
\end{align}
so that no new Feynman rules involving the $b'$ field have to be coded into FeynArts. 

The sector decomposition of $\Delta^{(1)}_{\omega \psi_i \bar \psi_i}$ can be seen from Fig.~\ref{fig:c_f_f}: 
$\Delta^{(1;1,0)}_{\omega \psi_i \bar \psi_i}$ is given by diagram $D1$, $\Delta^{(1;0,1)}_{\omega \psi_i \bar \psi_i}$ by diagram $D5$ while all the remaining diagrams contribute to
$\Delta^{(1;0,0)}_{\omega \psi_i \bar \psi_i}$.
In this latter sector diagram $D3$ cancels against 
diagram $D4$ since the $\sigma'$ and $X'$-propagators compensate exactly (see Eq.(\ref{scalar_mass_eigenstate_propagators})).
Also the propagators of $A''_\mu$ and $X''_\mu$ exactly cancel out among themselves, so in the combination of 
diagram $D6$ with diagram $D7$ 
only the piece of the $A'''_\mu$-propagator due to the incorporation of  $b'$-field survives.
Hence the contribution to $\Delta^{(1;0,0)}_{\omega \psi_i \bar \psi_i}$ reduces to the Goldstone exchange diagram $D2$ plus the terms
from the sum of diagrams $D6$ and $D7$.

Notice that the propagator of the gauge-invariant field $a'_\mu$ does not exhibit any pole in the longitudinal part, see Eq.(\ref{vector_mass_eigenstate_propagators}). The reshuffling of the unphysical poles in the gauge-Goldstone sector has been automatically taken into account by the gauge-invariant dynamical fields.
The price to pay is that for large momenta the propagator of the gauge-invariant field $a'_\mu$ falls off as a constant in its longitudinal component. Thus one expects contributions to $\Delta^{(1;0,1)}_{\omega \psi_i \bar \psi_j}$ that are not power-counting renormalizable. They are guaranteed to cancel against left-over terms from the sum of diagrams $D6$ and $D7$, however the different sectors need not respect the power-counting rules of the full amplitude.
The resulting expressions for the massive theory in the individual sectors are  reported in Appendix~\ref{app.sector.fermionic.amps}.  
We quote here the final result for the full breaking 
\begin{align}
    \Delta^{(1)}_{0;\omega \bar \psi_i(p_1) \psi_j(p_2) } & = 
    \Delta^{(1;0,0)}_{0;\omega \bar \psi_i \psi_j }
    + \Delta^{(1;0,1)}_{0;\omega \bar \psi_i \psi_j } +
    \Delta^{(1;1,0)}_{0;\omega \bar \psi_i \psi_j } 
    \nonumber \\
    &  
 =\frac{i e}{32\pi^2} (y_{L,i} - y_{R,i}) \Bigg\{
 2K^2 (\bar{\slashed p}_1+\bar{\slashed p}_2)(P_L-P_R) \nonumber \\
 & \qquad\qquad\qquad +\frac{5e^2}{3}\Big[
 4m_e y_{L,i}y_{R,i}(P_L-P_R)
 +(\bar{\slashed p}_1+\bar{\slashed p}_2)
  \big(y_{L,i}^{2}P_R-y_{R,i}^{2}P_L\big)
 \Big]
 \Bigg\},
\end{align}
\begin{align}
    \Delta^{(1)}_{1;\omega \bar \psi_i(p_1) \psi_j(p_2) } & = 
    \Delta^{(1;0,0)}_{1;\omega \bar \psi_i \psi_j }
    + \Delta^{(1;0,1)}_{1;\omega \bar \psi_i \psi_j } +
    \Delta^{(1;1,0)}_{1;\omega \bar \psi_i \psi_j } 
    \nonumber \\
    & = -\frac{i e^3}{6\pi^2}y_{L,i}y_{R,i}
 \left[
 (y_{R,i}\widehat{\slashed p}_1
 +y_{L,i}\widehat{\slashed p}_2)P_R+(y_{L,i}\widehat{\slashed p}_1
 +y_{R,i}\widehat{\slashed p}_2)P_L
 \right] \, .
\end{align}
Notice that no $1/M_A$-dependence is left over in the full breaking amplitude coefficients $\Delta^{(1)}_{0,1;\omega \bar \psi_i(p_1) \psi_j(p_2) }$. This is at variance with the $M_A$-dependence of the individual gauge-invariant sectors, as discussed in Appendix~\ref{app.sector.fermionic.amps}.

\subsection{Amplitudes with the insertion of one ghost, fermion-antifermion fields and scalars or vectors}

A similar analysis applies to amplitudes with an additional scalar or vector insertion.
In the gauge-invariant field formalism the presence of derivative interactions spoils power-counting in the intermediate stage in the individual 
$(\lh,\lap)$-sector. This in turn implies that higher power of momenta appear in the numerator. Moreover, in our computations we retain the full mass dependence, so the standard one-loop tensor reduction proves to be computationally demanding.

In the evaluation of the ST breaking one does not really needs the full exact amplitude.
It is in fact sufficient to compute the relevant UV and finite local parts. 

Both can be efficiently evaluated by using the so-called all massive tadpole (AMT) algorithm~\cite{Chetyrkin:1997fm,Lang:2021hnw}. The idea is to Taylor-expand amplitudes around zero momenta up to the relevant superficial UV degree $\delta$ of the particular diagram. If massless propagators are present, one recursively replaces them via the following exact identity
\begin{align}
\frac{1}{(k+p)^2} = \frac{1}{k^2-M^2} - \frac{p^2+ 2 kp + M^2}{k^2-M^2}\frac{1}{(k+p)^2} 
\end{align}
up to the required order.
In the above expression $k$ denotes the loop momentum and $M$ is an auxiliary scale. By performing the replacement and the Taylor expansion up to a sufficiently high order so that one captures all the UV divergent contributions, the resulting expressions provide the required UV and finite contributions. One has to evaluate expanded amplitudes that only contain massive tadpoles, while the momentum dependence is exactly given by the algebraic Taylor expansion.
No dependence on $M$ must appear in the results. This provides a useful consistency check of the computations.

The AMT algoritm has been implemented in Mathematica with the FeynCalc package. The preliminary version of the AMT subtraction implementation in Mathematica has been developed with the help of~\cite{OpenAIChatGPT2026}, before final revision by the Author.
All computations have been carried out in Mathematica/FeynCalc with deterministic code.

The expression of each individual $(\lh,\lap)$ sector are rather involved and reported in Appendix~\ref{app.sector.fermionic.amps}.
Here we give the result for the full amplitudes (sum over all $(\lh,\lap)$-sectors):

\begin{itemize}
\item $\Delta^{(1)}_{\omega(q) \bar \psi_i(p_1) \psi_j(p_2) \sigma(p_3)}$

By summing the sector results in Eqs.(\ref{UFFS4.20})-(\ref{UFFS4.00}) and 
by taking into account the relations in Eq.(\ref{UFFS4.relations}) one can check that all momentum contributions simplify and one gets the simple expressions
\begin{align}
  & \Delta^{(1)}_{0;\omega(q) \bar \psi_i(p_1) \psi_j(p_2) \sigma(p_3)} = -\frac{5ie^3K}{24\pi^2}\,
y_{L,i}y_{R,i}(y_{L,i}-y_{R,i})(P_R-P_L) \, , \nonumber \\  
  & \Delta^{(1)}_{1;\omega(q) \bar \psi_i(p_1) \psi_j(p_2) \sigma(p_3)} = 0 \, .
\end{align}

\item $\Delta^{(1)}_{\omega(q) \bar \psi_i(p_1) \psi_j(p_2) \chi(p_3)}$

Again by summing over all sectors as in Eqs.(\ref{UFFS2.11})-(\ref{UFFS2.00}) and by taking into account
Eq.(\ref{UFFS2.relations}) one sees that the three sectors’ pole coefficients cancel and the finite part is given by
\begin{align}
  & \Delta^{(1)}_{0;\omega(q) \bar \psi_i(p_1) \psi_j(p_2) \chi(p_3)} = \frac{5e^3K}{24\pi^2}\,
y_{L,i}y_{R,i}(y_{L,i}-y_{R,i})(P_R+P_L)
 \, , \nonumber \\  
  & \Delta^{(1)}_{1;\omega(q) \bar \psi_i(p_1) \psi_j(p_2) \chi(p_3)} = 0 \, .
\end{align}
\item $\Delta^{(1)}_{\omega(q) \bar \psi_i(p_1) \psi_j(p_2) A_\mu(p_3)}$

By inspecting Eqs.(\ref{UFFV2.11head})-(\ref{UFFV2.10}), we see that both the finite and pole coefficients cancel between sector $(1,0)$ and $(1,1)$. In the same way,
from Eqs.(\ref{UFFV2.01head}),(\ref{UFFV2.01}) and (\ref{UFFV2.00head}),(\ref{UFFV2.00}) we see that also the finite and pole coefficients cancel out between the sectors $(0,1)$ and $(0,0)$.
We end up with
\begin{align}
  & \Delta^{(1)}_{0;\omega(q) \bar \psi_i(p_1) \psi_j(p_2) A_\mu(p_3)} = 0 
 \, , \nonumber \\  
  & \Delta^{(1)}_{1;\omega(q) \bar \psi_i(p_1) \psi_j(p_2) A_\mu(p_3)} = 0 \, .
\end{align}
\end{itemize}

\subsection{External sources insertions in the fermionic ST breaking sector}

In the gauge-invariant formalism one also need to consider insertions of the external sources $\bar c^*$, $\bar c^*_\mu$ governing the dependence on the $X, X_\mu$-fields via Eqs.(\ref{X_equation}) and (\ref{Xmu_equation}).
One can easily see that by power-counting 
\begin{align}
       \Delta^{(1;\lh,\lap)}_{j;\omega \bar \psi_i(p_1) \psi_j(p_2) \bar c^*(p_3) } = 0 \, , \qquad j=0,1,
\end{align}
in all relevant sectors $(\lh, \lap)$, i.e. diagrams contributing to all gauge-invariant sectors for this amplitude
are separately superficially convergent by power-counting. This is because the interaction vertices for $\bar c^*$ are momentum-independent.

On the other hand, diagrams contributing to $\Delta^{(1;\lh,\lap)}_{j;\omega \bar \psi_i(p_1) \psi_j(p_2) \bar c^*_\mu(p_3)}$ are not in general superficially convergent by power-counting. By explicit computation one however finds that the ST breaking in each sector vanishes also for the Green's functions with a $\bar c^*_\mu$-insertion, i.e.
\begin{align}
       \Delta^{(1;\lh,\lap)}_{j;\omega \bar \psi_i(p_1) \psi_j(p_2) \bar c^*_\mu(p_3) } = 0 \, , \qquad j=0,1,
\end{align}
in all relevant sectors $(\lh, \lap)$.

%

\section{One-loop symmetry-restoring counterterms}\label{sec.cts}

In this Section we present an algorithm for the determination of a particular solution of the symmetry-restoring counter-terms once the breaking $\Delta^{(1)}$ is known.
The solution is not unique, since one can always add a linear combination of ${\cal S}_0$-invariants of dimension $\leq 4$ without spoiling the
validity of
\begin{align}
{\cal S}_0 (-\Xi^{(1)}) = \Delta^{(1)} \, .
\label{brst.invs}
\end{align}

We first notice that in the Landau gauge and for an Abelian gauge group the ghosts are free, so there are no radiative corrections to the antifields. Thus they cannot enter the ST breaking $\Delta^{(1)}$.

We then use integration by parts on $\Delta^{(1)}$ so that one ghost is left undifferentiated. Since in the Landau gauge and for an Abelian gauge group there is no dependence either on the anti-fields or the ghosts, $\Delta^{(1)}$ contains at most one undifferentiated $\omega$.

This in turn simplifies considerably the problem. 
In more general cases, one must deal with differentiated ghosts and make use of the well-known homotopy techniques in jet spaces~\cite{Barnich:2000zw}.

Eq.(\ref{brst.invs}) can be solved by expanding both sides in a complete basis of Lorentz covariant monomials of appropriate dimensions and ghost number.
The explicit expression for the breaking up to dimension $5$ are given in Appendix~\ref{app:cts}.

It should be emphasized that the inverse is not unique. One can in fact add a linear combination of BRST invariant 
terms of ghost number zero and dimension $\leq 4$ while preserving the breaking. Those terms have to be fixed by suitable normalization conditions.

The derivation of the BRST inverse of the ST breaking $\Delta^{(1;\lh,\lap)}$ for each independent
gauge-invariant sectors can be performed in a similar way and coded into a computer-algebra framework. The resulting functionals do not lend themselves to be presented in a compact analytical 
expression. This is due to the large amount of independent form factors present in the BMHV scheme, as a consequence of the decomposition of Lorentz covariants and Dirac matrices into a purely four-dimensional and an evanesent $D-4$ components. 
For this reason we limit ourselves to provide the 
explicit results for the symmetry-restoring counter-terms in the sector
of dimension $\leq 5$.

\section{Comparison with results in the standard formalism}\label{sec:comparison}

In order to compare with the results of Ref.~\cite{Ebert:2024xpy} we need first to go on shell
with the $X$ and $X_\mu$ Lagrange multipliers, sum over the sectors contributing to a given ST breaking and then take the massless limit $v \rightarrow 0$.

The elimination of the auxiliary fields $X,X_\mu$ can be formulated systematically by constructing the mapping between the theory with dynamical gauge-invariant fields and the standard theory 
where those fields have been eliminated via their equation of motions and those of their
Lagrange multiplier partners $X,X_\mu$ (target theory),
as explained in
Ref.~\cite{Binosi:2017ubk}. 

The $X$- and $X_\mu$-equations (\ref{X_equation}) and (\ref{Xmu_equation}) imply, 
for every loop order $j\geq 1$, that 
the dependence of the loop vertex functional on $X$ and $X_\mu$ can occur only through the shifted external sources
\begin{align}
\widetilde{\bar c}^{,*}
\equiv
\bar c^{*}+(\Box+m^{2})X,
\qquad
\widetilde{\bar c}^{,*\mu}
\equiv
\bar c^{*\mu}+\Sigma_{(0)}^{\mu\nu}X_{\nu}.
\label{repl.lagrange}
\end{align}
Thus, amplitudes containing external $X$- or $X_\mu$-legs are not independent amplitudes: they can be obtained algebraically by amplitudes with insertions of $\bar c^{*}$ or $\bar c^{*\mu}$, respectively
via the replacement in Eq.(\ref{repl.lagrange}).

To obtain the amplitudes of the target theory, one first uses the classical equations of motion of the Lagrange multipliers,
$$
\frac{\delta\Gamma^{(0)}}{\delta X}
=
(\Box+m^{2})
\bigl[h-{\mathcal H}(\phi)\bigr]=0,
\qquad
\frac{\delta\Gamma^{(0)}}{\delta X_\mu}
=
\Sigma_{(0)}^{\mu\nu}
\bigl[a_\nu-{\mathcal A}_\nu(\phi,A)\bigr]=0,
$$
where $\mathcal H$ and $\mathcal A$ are given by the FMS fields in Eq.(\ref{fms.fields}).

After discarding the zero modes of the homogenoues solutions, these equations give
$$
h={\mathcal H}(\phi),
\qquad
a_\mu={\mathcal A}_\mu(\phi,A).
$$
The remaining equations of motion, namely those of $h$ in Eq.(\ref{h_equation}) and $a_\mu$
in Eq.(\ref{amu_equation}), read respectively for $m=M_H$
$$
(\Box+m^{2})X
+\bar c^{*}=0,
\qquad
\Sigma_{(0)}^{\mu\nu}X_\nu+\bar c^{*\mu}=0.
$$
It follows from Eq.(\ref{repl.lagrange}) that, on the solutions of the auxiliary-field equations,
both sources $\widetilde{\bar c}^{,*}$ and $\widetilde{\bar c}^{,*\mu}$ are mapped to zero.

Therefore, 
the mapping from the gauge-invariant formulation to the target theory for the specific choice 
$m=M_H$ becomes particularly simple and 
can be summarized as
\begin{align}
\left.
\Gamma^{(j)}_{\mathrm{target}}
[\phi,A_\mu,\psi_i,\overline{\psi}_i]
=
\Gamma^{(j)}
\left[
\phi,A_\mu,\psi_i,\overline{\psi}_i;
\widetilde{\bar c}^{,*},
\widetilde{\bar c}^{,*\mu}
\right]
\right|_{
\substack{
\widetilde{\bar c}^{*}
=0\, ,
\widetilde{\bar c}^{*\mu}=0
}}
\label{target.theory}
\end{align}
with $h$ and $a_\mu$ replaced by $\mathcal H$ and $\mathcal A_\mu$. 
Functional differentiation of both sides of Eq.(\ref{target.theory}) with respect to the elementary fields and external sources generates the corresponding target-theory 1-PI amplitudes. 
Finally, the physical $S$-matrix is obtained from the target-theory Green functions by the usual LSZ reduction. 
The same prescription can be used to obtain the ST breaking $\Delta_{\mathrm{target}}^{(j)}$ from 
the ST breaking~$\Delta^{(j)}$.

We have checked our results against those of Ref.~\cite{Ebert:2024xpy}. In order to work out the matching, the following conventions must be used.
The massless limit $v \rightarrow 0$ has to be taken, since results of Ref.~\cite{Ebert:2024xpy} refer to a massless Abelian theory. The $\phi,\phi^\dagger$ fields of Ref.~\cite{Ebert:2024xpy} are related to $\sigma, \chi$ by
\begin{align}
    \phi = \frac{1}{\sqrt{2}} (\sigma + i \chi) \, , \qquad \phi^\dagger = \frac{1}{\sqrt{2}} ( \sigma - i \chi ) \, .
\end{align}
The coupling constants $G'$s in Ref.~\cite{Ebert:2024xpy}  are to be set to zero, while our $K$ differs by a factor $1/\sqrt{2}$ from the conventions of Ref.~\cite{Ebert:2024xpy}.
For open fermion chains, we remark that our amplitudes are reported in reversed order w.r.t. those of ~\cite{Ebert:2024xpy}, so one must apply the charge conjugation operator before comparison.

Once these steps are carried out, one finds exact agreements for all ST breaking amplitudes (in the massless limit) with the results of Ref.~\cite{Ebert:2024xpy}.

\section{Mass dependence via  extended Wess-Zumino consistency condition}\label{sec.wz}

It is interesting to note that the dependence of the
ST breaking $\Delta^{(n)}$ on the v.e.v. $v$ can
be constrained by extending the BRST differential $s$ in such a way that the v.e.v. $v$ is mapped into an anticommuting constant ghost $\Omega$, namely one defines
\begin{align}
s v = \Omega \, , \qquad s v = 0 \, .
\end{align}
The BRST transformation of the physical Higgs scalar $\sigma$ is correspondingly modified in such a way that
the combination $\phi_0 = \sigma + v$ still transforms as
\begin{align}
    s \phi_0 = e \omega \phi_0 \, ,
\end{align}
so that the classical action remains gauge invariant.
In Landau gauge no $\Omega$-dependent interactions arise, at variance with the $R_\xi$-gauge. In the latter case in fact the
gauge-fixing function ${\cal F}_\xi \equiv \partial A - \xi e v \chi$ depends on $v$, so that the gauge-fixing plus ghost part of the action contains a $\Omega$-dependent interaction vertex:
\begin{align}
    S_{\mbox{g.f + ghost}} = \int d^D x ~s \Big [ \bar \omega \Big ( \frac{1}{2\xi} b + {\cal F}_\xi \Big ) \Big ] \supset \int d^D x\, \xi \bar \omega e \Omega \chi \, .
\end{align}
This gives rise to non-trivial correlators with $\Omega$ external legs .

One can easily work out the transformation of $\sigma$:
\begin{align}
    s \phi_0 = s \sigma + s v = e  \omega (\sigma + v) \Rightarrow s \sigma = e  \omega (\sigma + v) - \Omega \, .
\end{align}
In Landau gauge the extended BRST differential ensures that the only dependence on the v.e.v. is via the combination
$\phi_0 = \sigma + v$. 

Let us now write down the one-loop Wess-Zumino consistency condition for the breaking $\Delta^{(1)}$ based on the extended BRST differential. 
We denote the linearized ST operator for the extended BRST differential by $\extS0$.
Again we use the fact that in Landau gauge and in the Abelian theory ghosts are free, so at one loop order all amplitudes involving the BRST antifields are zero. Thus one finds
\begin{align}
\extS0(\Delta^{(1)}) = \int d^Dx\,&\Bigg\{
\partial_\mu\omega\,
\frac{\delta \Delta^{(1)}}{\delta A_\mu}
+
b\,\frac{\delta \Delta^{(1)}}{\delta\bar\omega} + (-e\omega\chi - \Omega)
\frac{\delta \Delta^{(1)}}{\delta\sigma}
+
e \omega( v + \sigma)
\frac{\delta \Delta^{(1)}}{\delta\chi}
+ b \frac{\delta \Delta^{(1)}}{\delta \bar \omega}
\nonumber\\
&+\sum_{i=1}^{N_f} \Big [  
i e \omega (y_{L,i} P_L + y_{R,i} P_R) \psi_i\frac{\Delta^{(1)}}{\delta \psi_i} 
- i e \omega \bar \psi_i (y_{L,i} P_R + y_{R,i} P_L)  \frac{\Delta^{(1)}}{\delta \bar \psi_i} 
\Big ]  \nonumber \\
& + \Omega \frac{\partial \Delta^{(1)}}{\partial v} = 0 \, .
\label{1.loop.ext.sti}
\end{align}
By taking one derivative w.r.t $\omega$, one w.r.t $\Omega$ and any number of fields $\Phi_1,\dots,\Phi_n$ and then
setting the fields and external sources to zero one gets from Eq.(\ref{1.loop.ext.sti})
\begin{align}
    \frac{\partial}{\partial v} \frac{\delta^{(n+1)} \Delta^{(1)}}{\delta \omega(q) \delta \Phi_1(p_1) \dots \delta \Phi_n(p_n)} = \frac{\delta^{(n+2)} \Delta^{(1)}}{\delta \omega(q) \delta \sigma(0) \delta \Phi_1(p_1) \dots \delta \Phi_n(p_n)} \, .
    \label{vev.der}
\end{align}
This means that one can trade a derivative w.r.t $v$ with the insertion of an external $\sigma$-leg at zero momentum.
In particular, the above equation entails that the massless limit $v\rightarrow 0$ must be smooth: the ST breaking in the spontaneously broken phase must reproduce the ST breaking in the massless theory in the limit of vanishing v.e.v.

This result has been  checked on the explicit results for all relevant amplitudes. The formal proof is provided by Eq.(\ref{vev.der}).

A comment is in order here. Eq.(\ref{vev.der}) holds true in each $(\lh,\lap)$-sector.
The explicit results given in Appendix~\ref{app.sector.fermionic.amps} are obtained for a set of Feynman rules 
based on the condition in Eq.(\ref{m.mH}). This introduces an extra $v$-dependence that is not taken into account in Eq.(\ref{vev.der}). For full amplitudes the dependence on the parameter $m$ is known to cancel out~\cite{Binosi:2020unh} and consequently one can check that the explicit results for the full amplitudes do obey Eq.(\ref{vev.der}).
On the other hand, amplitudes in the individual $(\lh,\lap)$-sector receive an additional contribution arising from the fact that  the mass $M_h$ of the $h$-field is not an independent parameter, but it equals $M_h=m=\lambda v$, $\lambda$ being the Higgs coupling constant.

\section{Conclusions}\label{sec.conclusions}

We have evaluated the ST breaking and their symmetry-restoring counter-terms in an Abelian chiral gauge theory in Landau gauge in the BMHV scheme.
We have retained the full mass dependence. The gauge-invariant FMS formalism has been used in order to decompose the ST identities into independently invariant sectors. 
We have shown that the dependence of the ST breaking on the v.e.v. $v$ is controlled algebraically by the Wess-Zumino consistenty condition for an extended BRST differential mapping $v$ into a constant anticommuting ghost $\Omega$.
This in turn ensures that the massless limit is indeed smooth.
We have compared the complete ST breaking (sum over all $(\lh,\lap)$-sectors) with existing results in the literature and found complete agreement in the relevant massless limit. 
The gauge-invariant formalism requires more amplitudes to be evaluated than the standard one, yet use
of the all-massive-tadpole algorithm makes the computations easily accessible.
The results presented in this paper pave the way for applications to non-Abelian gauge group as well as to effective field theories (EFT). In the latter case, the gauge-invariant sector decomposition might prove particularly interesting, since it allows to control in a mathematically consistent way the independent contributions of physical gauge-invariant states insides loops, in particular those of massive Beyond-the-Standard-Model excitations that appear in various combinations together the Standard Mode fields. This means that one can grade the weight of individual loops in assessing their impact on the finite non-symmetric counterterms, to be added to regularized amplitudes in order to ensure the fulfillment of the EFT ST identities in precision calculations.
\appendix

\section*{Acknowledgments}

Useful discussions with D.~St\"ockinger and P.~K\"uhler are gratefully acknowledged.
\section{BRST differentials}

The BRST differential $s$ associated with U(1) gauge invariance is given by (we set $Y_S=1$)
\begin{flalign}
& s A_\mu
=\partial_\mu\omega,
\qquad
s\omega=0,
\qquad
s\bar\omega=b\, ,
\qquad
s b=0 \, ,
&& \nonumber\\
& s\phi
=ie \omega\,\phi \, ,
\qquad
s\phi^\dagger
=-ie \omega\,\phi^\dagger \, ,
\qquad
s\sigma
=-e \omega\,\chi \, ,
\qquad
s\chi
=e \omega\,(v+\sigma) \, ,
&& \nonumber\\
& s\psi_i
=ie\omega
\left(
y_{L,i}P_L+y_{R,i}P_R
\right)\psi_i \, ,
\qquad
s\overline{\psi}_i
=-ie\omega\,\overline{\psi}_i
\left(
y_{L,i}P_R+y_{R,i}P_L
\right) \, ,
&& \nonumber\\
& s h=0 \, , 
\qquad
s a_\mu=0 \, ,
\qquad
s X=s c=s\bar c=0 \, ,
\qquad
s X_\mu=s c_\mu=s\bar c_\mu=0 \, .
&& \label{U1.brst}
\end{flalign}
The gauge-invariant FMS fields $h, a_\mu$ in the last line are BRST-invariant by construction. The 
Lagrange multipliers $X,X_\mu$ and constraint ghosts $c, c_\mu$ and antighosts $\bar c, \bar c_\mu$ are also
invariant under $s$. However they are non-physical, since they form BRST doublets w.r.t. the 
constraint BRST differential $\s$, as can be seen from
\begin{flalign}
& \mathcal{s}\,\bar c
=
h-\frac{1}{v}
\left(
\phi^\dagger\phi-\frac{v^2}{2}
\right)\, ,
\qquad
\mathcal{s}\,X=c \, ,
\qquad
\mathcal{s}\,c=0 \, ,
\qquad
\mathcal{s}\,h=0 \, ,
\nonumber\\
& \mathcal{s}\,\bar c_\mu
=
a_\mu
-\frac{i}{ev^2}
\left[
2\phi^\dagger D_\mu\phi
-\partial_\mu\left(\phi^\dagger\phi\right)
\right] \, ,
\qquad
\mathcal{s}\,X_\mu=c_\mu \, ,
\qquad
\mathcal{s}\,c_\mu=0 \, ,
\qquad
\mathcal{s}\,a_\mu=0 \, . 
\label{constraint_BRST}
\end{flalign}
One says that $u,t$ are BRST doublets if $su=t\, , st =0$.
In this case one can prove~\cite{Gomis:1994he} that the cohomology of $s$ in independent of $u,t$.

The action of $\s$ vanishes on the other fields.

\section{Functional identities of the model}
\label{app.funct.ids}

The $D$-dimensional vertex functional obeys
the following functional identities:
\begin{itemize}
\item the $b$-equation
    \begin{equation}
\frac{\delta\Gamma}{\delta b}
=
-\partial^\mu A_\mu \, .
\label{b_equation}
\end{equation}
This equation implies that there is no dependence on the Nakanish-Lautrup field $b$ at the quantized level:
\begin{equation}
\frac{\delta\Gamma^{(j)}}{\delta b}
= 0 \, , \qquad j \geq 1 \, .
\label{b_equation.j}
\end{equation}
\item the U(1) ghost and antighost equation
\begin{equation}
\frac{\delta\Gamma^{(0)}}{\delta\bar\omega}
=
\Box\omega \, , \qquad\frac{\delta\Gamma^{(0)}}{\delta\omega}
=
-\Box\bar\omega \,.
\label{U1_ghost_equation}
\end{equation}
These equations hold in the Landau gauge and for an Abelian gauge group and imply that the ghost and antighost fields are free. 
\item the $X$-equation
\begin{equation}
\frac{\delta\Gamma}{\delta X}
\equiv
(\Box+m^2)
\frac{\delta\Gamma}{\delta\bar c^*}.
\label{X_equation}
\end{equation}
The above equation implies that $\G$ depends on
$X$ only via the combination
$$\widetilde{\bar c}^* = {\bar c}^* + (\square + m^2) X \, .$$
\item the $X_\mu$-equation
\begin{equation}
\frac{\delta\Gamma}{\delta X_\mu}
=
\Sigma^{\mu\nu}_{(0)}
\frac{\delta\Gamma}{\delta\bar c^{*\nu}}.
\label{Xmu_equation}
\end{equation}
As for the $X$-equation, this equation implies that  $\G$ depends on
$X_\mu$ only via the combination
$$\widetilde{\bar c}^{*\mu} = {\bar c}^{*\mu} + \Sigma^{\mu\nu}_{(0)} X_\nu \, .$$
\item the ghost-antighost equations for the constraint scalar and vector ghost and antighost
fields
\begin{equation}
\frac{\delta\Gamma}{\delta\bar c}
=
-(\Box+m^2)c,
\qquad
\frac{\delta\Gamma}{\delta c}
=
(\Box+m^2)\bar c.
\label{scalar_constraint_ghost_equations}
\end{equation}
\begin{equation}
\frac{\delta\Gamma}{\delta\bar c_\mu}
=
-\Sigma^{\mu\nu}_{(\xi)}c_\nu,
\qquad
\frac{\delta\Gamma}{\delta c_\mu}
=
\Sigma^{\mu\nu}_{(\xi)}\bar c_\nu.
\label{vector_constraint_ghost_equations}
\end{equation}
The above equations imply that $\bar c, c$ and $\bar c_\mu, c_\mu$ are free.
\item the scalar and vector constraint ST identities
\begin{equation}
{\cal S}_{\mathrm{con},h}\!\left(\Gamma\right)
\equiv
\int d^4x\,
\left[
c\,\frac{\delta\Gamma}{\delta X}
+
\frac{\delta\Gamma}{\delta\bar c^*}
\frac{\delta\Gamma}{\delta\bar c}
\right]
=0 \, ,
\label{scalar_constraint_ST}
\end{equation}
\begin{equation}
{\cal S}_{\mathrm{con},a}\!\left( \Gamma \right)
\equiv
\int d^Dx\,
\left[
c_\mu\frac{\delta\Gamma}{\delta X_\mu}
+
\frac{\delta\Gamma}{\delta\bar c_\mu^*}
\frac{\delta\Gamma}{\delta\bar c_\mu}
\right]
=0.
\label{vector_constraint_ST}
\end{equation}
By using the first of Eqs.(\ref{scalar_constraint_ghost_equations}), (\ref{vector_constraint_ghost_equations})
one sees that Eqs.(\ref{scalar_constraint_ST}),(\ref{vector_constraint_ST}) are equivalent to 
Eq.(\ref{X_equation}) and Eq.(\ref{Xmu_equation}) respectively.
\item the $h$ and $a_\mu$-equations
\begin{equation}
\frac{\delta\Gamma}{\delta h}
=
(\Box+m^2)X
+\bar c^* \, ,
\label{h_equation}
\end{equation}
\begin{equation}
\frac{\delta\Gamma}{\delta a_\mu}
=
\Sigma^{\mu\nu}_{(0)}X_\nu
+\bar c^{*\mu}.
\label{amu_equation}
\end{equation}
The above equations imply
\begin{equation}
\frac{\delta\Gamma^{(j)}}{\delta h} = \frac{\delta\Gamma^{(j)}}{\delta a_\mu} = 0 \, , \qquad j \geq 1 \, .
\label{g.inv.indep.symm.basis}
\end{equation}

\end{itemize}

\section{Propagators}\label{app.propagators}

In what follows $M_H$ denotes the physical Higgs mass, while
\begin{equation}
M_A=e v
\end{equation}
is the gauge-boson mass. 
We define the transverse and longitudinal projectors as
\begin{equation}
T_{\mu\nu}(p)
=
g_{\mu\nu}-\frac{p_\mu p_\nu}{p^2},
\qquad
L_{\mu\nu}(p)
=
\frac{p_\mu p_\nu}{p^2},
\end{equation}
and use the convention
\begin{equation}
\Delta_{\Phi_1\Phi_2}(p)
\equiv
\left\langle T
\Phi_1(-p)\Phi_2(p)
\right\rangle 
\end{equation}
for generic fields $\Phi_1$, $\Phi_2$ of incoming momentum $-p$ and $p$ 
respectively.

\subsection{Scalar sector}

The symmetric scalar basis is given by
\begin{equation}
\left(X,\sigma,h\right).
\end{equation}
The quadratic part of the scalar action is diagonalized through the
field redefinitions
\begin{equation}
X=-X'-h,
\qquad
\sigma=\sigma'+X'+h.
\label{scalar_diagonalization}
\end{equation}
Equivalently, the mass eigenstates are expressed in terms of the
symmetric-basis fields as
\begin{equation}
X'=-X-h,
\qquad
\sigma'=\sigma+X.
\end{equation}
The scalar mass-eigenstate basis is therefore
\begin{equation}
\left(\sigma',X',h\right).
\end{equation}

\subsubsection{Mass-eigenstate propagators}

The non-vanishing scalar propagators in the mass-eigenstate basis are (for $m=M_H$)
\begin{equation}
\Delta_{\sigma'\sigma'}(p)
=
-\Delta_{X'X'}(p)
=
\frac{i}{p^2-M_H^2},
\qquad
\Delta_{hh}(p)
=
\frac{i}{p^2-M_H^2}.
\label{scalar_mass_eigenstate_propagators}
\end{equation}

Thus, with the choice $m=M_H$ $\sigma'$, $X'$ and $h$
have the  pole at the same value $p^2=M_H^2$. Notice  the
relative minus sign between $\Delta_{\sigma'\sigma'}$ and
$\Delta_{X'X'}$ that ensures the mutual cancellations
of the unphysical modes described by $\sigma'$ and $X'$.

\subsubsection{Symmetric-basis propagators}

In the symmetric basis $(X,\sigma,h)$, the nonvanishing propagators are for $m=M_H$
\begin{equation}
\Delta_{XX}(p)
=
\Delta_{X\sigma}(p)
=
0,
\label{scalar_symmetric_propagators_1}
\end{equation}
and
\begin{equation}
\Delta_{hh}(p)
=
\Delta_{\sigma h}(p)
=
\Delta_{\sigma\sigma}(p)
=
-\Delta_{Xh}(p)
=
\frac{i}{p^2-M_H^2}.
\label{scalar_symmetric_propagators_2}
\end{equation}

\subsection{Vector sector in the Landau gauge}

The symmetric vector basis is
\begin{equation}
\left(A_\mu,\chi,b,X_\mu,a_\mu\right).
\end{equation}
The quadratic part of the action is diagonalized through a sequence of
local linear field redefinitions. The final mass eigenstates, expressed
in terms of the symmetric-basis fields, are
\begin{equation}
b'=b-M_A\chi,
\label{b_mass_eigenstate}
\end{equation}
\begin{equation}
A_\mu''
=
A_\mu-X_\mu
+\frac{1}{M_A^2}\partial_\mu b
-\frac{1}{M_A}\partial_\mu\chi,
\label{A_mass_eigenstate}
\end{equation}
and
\begin{equation}
X_\mu''
=
X_\mu-a_\mu
-\frac{1}{M_A^2}\partial_\mu b,
\qquad
a_\mu'
=
a_\mu+\frac{1}{M_A^2}\partial_\mu b.
\label{Xa_mass_eigenstates}
\end{equation}
The vector mass-eigenstate basis is therefore
\begin{equation}
\left(A_\mu'',X_\mu'',a_\mu',b',\chi\right).
\end{equation}
Notice that the combination $a'_\mu$ is also gauge-invariant.

\subsubsection{Mass-eigenstate propagators}

The diagonal vector propagators in the Landau gauge are
\begin{align}
\Delta_{A_\mu''A_\nu''}(p)
&=
\Delta_{a_\mu'a_\nu'}(p)
=
-\Delta_{X_\mu''X_\nu''}(p)
\nonumber\\
&=
\frac{i}{-p^2+M_A^2}\,T_{\mu\nu}(p)
+\frac{i}{M_A^2}\,L_{\mu\nu}(p).
\label{vector_mass_eigenstate_propagators}
\end{align}
The remaining nonvanishing propagators are
\begin{equation}
\Delta_{b'b'}(p)
=
-\frac{iM_A^2}{p^2},
\qquad
\Delta_{\chi\chi}(p)
=
\frac{i}{p^2}.
\label{bchi_mass_eigenstate_propagators}
\end{equation}
The physical massive pole is contained in the transverse part of the
$a_\mu'$ propagator. Its longitudinal part is momentum independent and
therefore does not describe an additional propagating pole. The fields
$A_\mu''$, $X_\mu''$, $b'$, and $\chi$ belong to the unphysical sector.

\subsubsection{Symmetric-basis propagators}

In the symmetric basis, the nonvanishing propagators are
\begin{equation}
\Delta_{A_\mu A_\nu}(p)
=
\frac{i}{-p^2+M_A^2}\,T_{\mu\nu}(p),
\label{AA_symmetric_propagator}
\end{equation}
\begin{equation}
\Delta_{A_\mu b}(p)
=
-\frac{p_\mu}{p^2},
\label{Ab_symmetric_propagator}
\end{equation}
and
\begin{equation}
\Delta_{A_\mu a_\nu}(p)
=
\frac{i}{-p^2+M_A^2}\,T_{\mu\nu}(p).
\label{Aa_symmetric_propagator}
\end{equation}
The propagators involving the gauge-invariant vector field and the
vector Lagrange multiplier are
\begin{align}
\Delta_{X_\mu a_\nu}(p)
&=
\frac{i}{-p^2+M_A^2}\,T_{\mu\nu}(p)
+\frac{i}{M_A^2}\,L_{\mu\nu}(p),
\\
\Delta_{a_\mu a_\nu}(p)
&=
\frac{i}{-p^2+M_A^2}\,T_{\mu\nu}(p)
+\frac{i}{M_A^2}\,L_{\mu\nu}(p).
\label{Xa_aa_symmetric_propagators}
\end{align}
The remaining non-vanishing mixed propagators are
\begin{equation}
\Delta_{a_\mu\chi}(p)
=
\frac{p_\mu}{M_Ap^2},
\qquad
\Delta_{b\chi}(p)
=
\frac{iM_A}{p^2},
\qquad
\Delta_{\chi\chi}(p)
=
\frac{i}{p^2}.
\label{mixed_symmetric_propagators}
\end{equation}
All the remaining propagators 
vanish. In particular,
\begin{equation}
\Delta_{X_\mu X_\nu}(p)
=
\Delta_{A_\mu X_\nu}(p)
=0 \, .
\label{vanishing_symmetric_propagators}
\end{equation}

The symmetric basis makes the functional identities and the
auxiliary-field hierarchy transparent, whereas the mass-eigenstate
basis exposes the physical gauge-invariant propagators $h$ and
$a_\mu'$. It is in the latter basis that 1-PI amplitudes can be graded
according to the numbers of internal $h$- and $a_\mu'$-lines.

The UV dimensions of the fields in the mass eigenstate basis 
and external sources in the Landau gauge are summarized in the following table:
\begin{table}[htbp]
\begin{center}
\begin{tabular}{c|c}
\hline
Fields or external sources & $d_{\mathrm{UV}}$ \\
\hline
$A_\mu^{\prime\prime}$, $a^\prime_\mu$, $X^{\prime \prime}_\mu$, 
$b^\prime$, $h$,
$\sigma'$, $X'$, $\chi$
& $1$ \\[1mm]

$\omega$, $\bar{\omega}$, $c$, $\bar{c}$,
$c_\mu$, $\bar{c}_\mu$
& $1$ \\[1mm]

$\psi_i$, $\overline{\psi}_i$
& $\frac{3}{2}$ \\[1mm]

$\sigma^*$, $\chi^*$, $\bar{c}^*$
& $2$ \\[1mm]

$\bar{c}^{*\mu}$
& $1$ \\[1mm]

$\eta_i$, $\bar{\eta}_i$
& $\frac{3}{2}$ \\
\hline
\end{tabular}
\end{center}
\caption{UV dimensions of the  fields and external sources in
the Landau gauge.}
\label{tab:uv_dimensions}
\end{table}

\section{One-loop sector-decomposition coefficients of fermionic amplitudes}\label{app.sector.fermionic.amps}

Let us decompose the momenta  into their four dimensional and $D-4$ components as
\begin{align}
 p_j =\bar p_j+\widehat p_j,
 \qquad
 s_{jk}\equiv \bar p_j\!\cdot\!\bar p_k,
 \qquad
 \widehat{s}_{jk}\equiv \widehat p_j\!\cdot\!\widehat p_k.
\end{align}
Consequently,
\begin{align}
 p_j \!\cdot p_k =s_{jk}+\widehat{s}_{jk}.
\end{align}

\subsection{Ghost-fermion-fermion}
We start from the ghost-fermion-fermion breaking.
\subsubsection{$(0,1)$-sector}
\begin{align}
\Delta^{(1;0,1)}_{0;\omega(q) \bar\psi_i(p_1) \psi_j(p_2) } = \delta_{ij} {}&\frac{i e^3}{192\pi^2M_A^2} (y_{L,i}-y_{R,i})
 \Big\{
 2m_e y_{L,i}y_{R,i}X_0(P_L-P_R)
 \nonumber\\
 &\quad
 +y_{L,i}^{2}
   \big(X_1\bar{\slashed p}_1+X_2\bar{\slashed p}_2\big)P_R
 -y_{R,i}^{2}
   \big(X_1\bar{\slashed p}_1+X_2\bar{\slashed p}_2\big)P_L
 \Big\} \, ,
\end{align}
where the three kinematic polynomials read
\begin{align}
 X_0&=20M_A^2+8m_e^2-s_{11}-3s_{12}-s_{22},\\
 X_1&=10M_A^2+4m_e^2-2s_{11}-3s_{12}+s_{22},\\
 X_2&=10M_A^2+4m_e^2+s_{11}-3s_{12}-2s_{22}.
\end{align}
\begin{align}
    \Delta^{(1;0,1)}_{1;\omega(q) \bar\psi_i(p_1) \psi_j(p_2) } = \delta_{ij} {}
    \frac{i e^3}{96\pi^2M_A^2}
 \sum_{\iota\in\{R,L\}}\Bigg[ &
 a_{\iota}^{(1)}P_\iota
 +\sum_{j=1}^{2}b_{j\iota}^{(1)}\bar{\slashed p}_jP_\iota
+\sum_{j=1}^{2}c_{j\iota}^{(1)}\widehat{\slashed p}_jP_\iota
 \nonumber \\
 &+\sum_{j,k=1}^{2}d_{jk,\iota}^{(1)}
   \bar{\slashed p}_j\widehat{\slashed p}_kP_\iota
 +e_{\iota}^{(1)}\widehat{\slashed p}_1\widehat{\slashed p}_2P_\iota
 \nonumber 
 \\
 &+\sum_{k=1}^{2}f_{k,\iota}^{(1)}
   \bar{\slashed p}_1\bar{\slashed p}_2
   \widehat{\slashed p}_kP_\iota
 +\sum_{j=1}^{2}g_{j,\iota}^{(1)}
   \bar{\slashed p}_j\widehat{\slashed p}_1
   \widehat{\slashed p}_2P_\iota
 \Bigg] \, ,
 \label{Delta_10_omega_psi_barpsi}
\end{align}
where the form factor coefficients are defined as follows.
We start from the projector coefficients, given by
\begin{align}
 a_R^{(1)}={}&-4m_e y_{L,i}y_{R,i}
 \Big[y_{L,i}(\widehat{s}_{11}+\widehat{s}_{12})
      -y_{R,i}(\widehat{s}_{12}+\widehat{s}_{22})\Big],\\
 a_L^{(1)}={}&+4m_e y_{L,i}y_{R,i}
 \Big[y_{L,i}(\widehat{s}_{12}+\widehat{s}_{22})
      -y_{R,i}(\widehat{s}_{11}+\widehat{s}_{12})\Big].
      \label{a.ff}
\end{align}
The four dimensional single-slash coefficients
in the first line of Eq.(\ref{Delta_10_omega_psi_barpsi})
are
\begin{align}
 b_{1R}^{(1)}={}&-y_{L,i}^{2}
 \Big[(y_{L,i}+y_{R,i})\widehat{s}_{11}
 +(y_{L,i}+6y_{R,i})(\widehat{s}_{12}+\widehat{s}_{22})\Big],\\
 b_{1L}^{(1)}={}&-y_{R,i}^{2}
 \Big[(y_{L,i}+y_{R,i})\widehat{s}_{11}
 +(6y_{L,i}+y_{R,i})(\widehat{s}_{12}+\widehat{s}_{22})\Big],\\
 b_{2R}^{(1)}={}&-y_{L,i}^{2}
 \Big[(y_{L,i}+6y_{R,i})(\widehat{s}_{11}+\widehat{s}_{12})
 +(y_{L,i}+y_{R,i})\widehat{s}_{22}\Big],\\
 b_{2L}^{(1)}={}&-y_{R,i}^{2}
 \Big[(6y_{L,i}+y_{R,i})(\widehat{s}_{11}+\widehat{s}_{12})
 +(y_{L,i}+y_{R,i})\widehat{s}_{22}\Big].
 \label{b.ff}
\end{align}

The evanescent slash coefficients in the
first line of Eq.(\ref{Delta_10_omega_psi_barpsi})
can be expressed via the master polynomial
\begin{align}
 \mathcal E_{jk}(x,y)={}&-160yM_A^2+40m_e^2(x-2y)
 +(10x+40y)s_{jj}-(20x+10y)s_{jk}
 \nonumber\\
 &+(15x+35y)s_{kk}+20y\widehat{s}_{jj}.
\end{align}
Then
\begin{align}
 c_{1R}^{(1)}&=\frac{y_{L,i}y_{R,i}}{10}
 \mathcal E_{12}(y_{L,i},y_{R,i}),
&
 c_{1L}^{(1)}&=\frac{y_{L,i}y_{R,i}}{10}
 \mathcal E_{12}(y_{R,i},y_{L,i}),\\
 c_{2R}^{(1)}&=\frac{y_{L,i}y_{R,i}}{10}
 \mathcal E_{21}(y_{R,i},y_{L,i}),
&
 c_{2L}^{(1)}&=\frac{y_{L,i}y_{R,i}}{10}
 \mathcal E_{21}(y_{L,i},y_{R,i}).
  \label{c.ff}
\end{align}
The mixed four-dimensional-evanescent slash form factor coefficients are
\begin{align}
 d_{11,R}^{(1)}=d_{22,R}^{(1)}
 &=-2m_e y_{L,i}^{2}(2y_{L,i}+y_{R,i}),\\
 d_{11,L}^{(1)}=d_{22,L}^{(1)}
 &=-2m_e y_{R,i}^{2}(y_{L,i}+2y_{R,i}),\\
 d_{12,R}^{(1)}=d_{21,R}^{(1)}
 &=+m_e y_{L,i}^{2}(y_{L,i}-7y_{R,i}),\\
 d_{12,L}^{(1)}=d_{21,L}^{(1)}
 &=-m_e y_{R,i}^{2}(7y_{L,i}-y_{R,i}).
 \label{d.ff}
\end{align}
The two-evanescent slash form factor coefficients are
\begin{align}
 e_R^{(1)}&=+4m_e y_{L,i}y_{R,i}(y_{L,i}-y_{R,i}),
&
 e_L^{(1)}&=-4m_e y_{L,i}y_{R,i}(y_{L,i}-y_{R,i}) \, ,
 \label{e.ff}
\end{align}
while the mixed two four-dimensional slashes and one evanescent slash form factor coefficients can be written as
\begin{align}
 f_{1,R}^{(1)}=f_{2,L}^{(1)}
 &=\frac{3}{2}y_{L,i}y_{R,i}(3y_{L,i}+5y_{R,i}),\\
 f_{1,L}^{(1)}=f_{2,R}^{(1)}
 &=\frac{3}{2}y_{L,i}y_{R,i}(5y_{L,i}+3y_{R,i}).
 \label{f.ff}
\end{align}
Finally, the one four-dimensional and two evanescent slashes form factor coefficients read
\begin{align}
 g_{1,R}^{(1)}=g_{2,R}^{(1)}&=-2y_{L,i}^{3},\\
 g_{1,L}^{(1)}=g_{2,L}^{(1)}&=-2y_{R,i}^{3}.
 \label{g.ff}
\end{align}
Notice that there is no smooth limit for $M_A$ going to zero.

\subsubsection{$(1,0)$-sector}
%
%
%
%
One finds
\begin{align}
& \Delta^{(1;1,0)}_{0;\omega(q) \bar\psi_i(p_1) \psi_j(p_2) } = \delta_{ij} 
 {}\frac{i e K^2}{32\pi^2} (y_{L,i}-y_{R,i})
 \Big\{
 2m_e \big(P_L-P_R\big)
 +\big(\bar{\slashed p}_1+\bar{\slashed p}_2\big)
  (P_L-P_R)
 \Big\} \, , \\
& \Delta^{(1;1,0)}_{1;\omega(q) \bar\psi_i(p_1) \psi_j(p_2) } = -\delta_{ij} {} 
 \frac{i e K^2}{16\pi^2}
\Big[
 \big(
 y_{L,i}\widehat{\slashed p}_1
 +y_{R,i}\widehat{\slashed p}_2
 \big)P_R
  +\big(
 y_{R,i}\widehat{\slashed p}_1
 +y_{L,i}\widehat{\slashed p}_2
 \big)P_L
 \Big].
\end{align}
\subsubsection{$(0,0)$-sector}

The sector $(0,0)$ is spanned by diagrams $D2$, $D6$ and $D7$ in Figure~\ref{fig:c_f_f}.

We separately give the expression of 
the diagram $D2$ with the Goldstone exchange
and the sum of diagrams $D6$ and $D7$:
\begin{itemize}
    \item diagram $D2$
 \begin{align}
    \Delta^{(1;0,0)}_{0,D2;\omega(q) \bar\psi_i(p_1) \psi_j(p_2) } = &-\delta_{ij} {}
    \frac{i e K^2}{32\pi^2} (y_{L;i}-y_{R;i}) \nonumber \\
    & \qquad \qquad \Big\{
 2m_e\big( P_L - P_R \big)
 -\big(\bar{\slashed p}_1+\bar{\slashed p}_2\big)
  (P_L-P_R)
 \Big\} \, , 
 \\
\Delta^{(1;0,0)}_{1,D2;\omega(q) \bar\psi_i(p_1) \psi_j(p_2) } = & \delta_{ij} 
\frac{i e K^2}{16\pi^2} \Big [
\big(
 y_{L,i}\widehat{\slashed p}_1
 +y_{R,i}\widehat{\slashed p}_2
 \big)P_R
 +\big(
 y_{R,i}\widehat{\slashed p}_1 +y_{L,i}\widehat{\slashed p}_2
 \big)P_L
 \Big].
\end{align}
Notice the relations
\begin{align}
&
\Delta^{(1;1,0)}_{0;\omega(q) \bar\psi_i(p_1) \psi_j(p_2) } + \Delta^{(1;0,0)}_{0,D2;\omega(q) \bar\psi_i(p_1) \psi_j(p_2) } = \delta_{ij} \frac{i e K^2}{16\pi^2}
\left(y_{L,i}-y_{R,i}\right)
\left(\bar{\slashed p}_1+\bar{\slashed p}_2\right)
\left(P_L-P_R\right)\, ,
\nonumber \\
& 
\Delta^{(1;1,0)}_{1;\omega(q) \bar\psi_i(p_1) \psi_j(p_2) } + \Delta^{(1;0,0)}_{1,D2;\omega(q) \bar\psi_i(p_1) \psi_j(p_2) } = 0 \, .
\end{align}
\item diagrams $D6$ and $D7$
\begin{align}
    \Delta^{(1;0,0)}_{0,D6+D7;\omega(q) \bar\psi_i(p_1) \psi_j(p_2) } = &-\delta_{ij} {}
    \frac{i e^3}{192\pi^2M_A^2} (y_{L,i}-y_{R,i})
 \Big\{
 2m_e y_{L,i}y_{R,i}\widetilde{X_0}(P_L-P_R)
 \nonumber\\
 &\quad
 +y_{L,i}^{2}
   \big(\widetilde{X}_1\bar{\slashed p}_1+\widetilde{X}_2\bar{\slashed p}_2\big)P_R
 -y_{R,i}^{2}
   \big(\widetilde{X}_1\bar{\slashed p}_1+\widetilde{X}_2\bar{\slashed p}_2\big)P_L
 \Big\} \, ,
\end{align}
where we have set
\begin{align}
\widetilde{X}_0 = X_0 - 20 M_A^2\, , \qquad \widetilde{X}_i  = X_i - 10 M_A^2\, , ~~~i=1,2 \, .
\end{align}
Moreover
\begin{align}
    \Delta^{(1;0,0)}_{1,D6+D7;\omega(q) \bar\psi_i(p_1) \psi_j(p_2) } = -\delta_{ij} {}
    \frac{i e^3}{96\pi^2M_A^2}
 \sum_{\iota\in\{R,L\}}\Bigg[ &
 a_{\iota}^{(1)}P_\iota
 +\sum_{j=1}^{2}b_{j\iota}^{(1)}\bar{\slashed p}_jP_\iota
+\sum_{j=1}^{2}\tilde{c}_{j\iota}^{(1)}\widehat{\slashed p}_jP_\iota
 \nonumber \\
 &+\sum_{j,k=1}^{2}d_{jk,\iota}^{(1)}
   \bar{\slashed p}_j\widehat{\slashed p}_kP_\iota
 +e_{\iota}^{(1)}\widehat{\slashed p}_1\widehat{\slashed p}_2P_\iota
 \nonumber \\
 &+\sum_{k=1}^{2}f_{k,\iota}^{(1)}
   \bar{\slashed p}_1\bar{\slashed p}_2
   \widehat{\slashed p}_kP_\iota
 +\sum_{j=1}^{2}g_{j,\iota}^{(1)}
   \bar{\slashed p}_j\widehat{\slashed p}_1
   \widehat{\slashed p}_2P_\iota
 \Bigg] \, ,
 \label{Delta_00_1_omega_psi_barpsi}
\end{align}
where the form factors are given by Eqs.(\ref{a.ff}), (\ref{b.ff}), (\ref{d.ff}), (\ref{e.ff}), (\ref{f.ff}), (\ref{g.ff}) and $\tilde c_{j\iota}$  are obtained from Eq.(\ref{c.ff}) after the replacement
$$ \mathcal E_{ij}(y_{L,i},y_{R,i}) \rightarrow \widetilde{\mathcal{E}}_{ij}(y_{L,i},y_{R,i})$$
where the tilded master polynomial is defined by
$$ \mathcal E_{jk}(x,y)
 =\widetilde{\mathcal E}_{jk}(x,y)-160yM_A^2. $$
We notice that all the momentum-dependent and mass-dependent terms proportional to
$1/M_A^2$  cancel between the finite coefficients of amplitude $D5$ and amplitudes $D6,D7$.  The result is
\begin{align}
\Delta^{(1;0,1)}_{0;\omega(q) \bar\psi_i(p_1) \psi_j(p_2) } &+
\Delta^{(1;0,0)}_{0,D6+D7;\omega(q) \bar\psi_i(p_1) \psi_j(p_2) } 
 ={}
 \nonumber \\
& \frac{5i e^3}{96\pi^2}\delta_{ij}(y_{L,i}-y_{R,i})
 \Big\{
 4m_e y_{L,i}y_{R,i}(P_L-P_R)
 +(\bar{\slashed p}_1+\bar{\slashed p}_2)
 \big(y_{L,i}^{2}P_R-y_{R,i}^{2}P_L\big)
 \Big\}.
\end{align}
In a similar way, by taking into account the $M_A$-dependence of the $\tilde{c}$'s coefficients
one sees that no $1/M_A$-term is left in the sum of the divergent coefficients, yielding
\begin{align}
\Delta^{(1;0,1)}_{1;\omega(q) \bar\psi_i(p_1) \psi_j(p_2) } &+
\Delta^{(1;0,0)}_{1,D6+D7;\omega(q) \bar\psi_i(p_1) \psi_j(p_2) } 
\nonumber \\
 =&{} -\frac{i e^3}{6\pi^2}y_{L,i}y_{R,i}\Big[
 \big(y_{R,i}\widehat{\slashed p}_1
     +y_{L,i}\widehat{\slashed p}_2\big)P_R
+\big(y_{L,i}\widehat{\slashed p}_1
       +y_{R,i}\widehat{\slashed p}_2\big)P_L
 \Big]. 
\end{align}
\end{itemize}
\subsection{Ghost-fermion-fermion-scalars}

In the following formulas we omit the common factor $\delta_{ij}$. One finds

\subsubsection{$\Delta^{(1)}_{\omega(q) \bar \psi_i(p_1) \psi_j(p_2) \sigma(p_3)}$}

The relevant contributions are
\begin{itemize}
\item $(2,0)$-sector
\begin{align}
    \Delta^{(1;2,0)}_{0;\omega(q) \bar \psi_i(p_1) \psi_j(p_2) \sigma(p_3)} & =
    -\frac{i e K^2}{24\pi^2 v}
\left(y_{L,i}-y_{R,i}\right)
\left(2Kv+\bar{\slashed p}_1+\bar{\slashed p}_2\right)
\left(P_R-P_L\right) \, , \\
 \Delta^{(1;2,0)}_{1;\omega(q) \bar \psi_i(p_1) \psi_j(p_2) \sigma(p_3)} & =
 -\frac{i e K^2}{6\pi^2 v}
\left[
\left(y_{L,i}\widehat{\slashed p}_1
+y_{R,i}\widehat{\slashed p}_2\right)P_R
+
\left(y_{R,i}\widehat{\slashed p}_1
+y_{L,i}\widehat{\slashed p}_2\right)P_L
\right] \, .
\label{UFFS4.20}
\end{align}
\item $(1,1)$-sector
\begin{align}
    \Delta^{(1;1,1)}_{0;\omega(q) \bar \psi_i(p_1) \psi_j(p_2) \sigma(p_3)} & = 0 \, , \\
    \Delta^{(1;1,1)}_{1;\omega(q) \bar \psi_i(p_1) \psi_j(p_2) \sigma(p_3)} & = 0 \, .
    \label{UFFS4.11}
\end{align}
\item $(0,2)$-sector
\begin{align}
    \Delta^{(1;0,2)}_{0;\omega(q) \bar \psi_i(p_1) \psi_j(p_2) \sigma(p_3)} & = \frac{i e(y_{L,i}-y_{R,i})}{960\pi^2v^3}
\left[
V^{(0,2)} \left(y_{L,i}^{2}P_R-y_{R,i}^{2}P_L\right) \right . \nonumber \\
& 
\left .
-4Kv\,y_{L,i}y_{R,i}F^{(0,2)}(P_R-P_L)
\right] \, , \\
    \Delta^{(1;0,2)}_{1;\omega(q) \bar \psi_i(p_1) \psi_j(p_2) \sigma(p_3)} & =  
    -\frac{ie}{480\pi^2v^3}
\left[
\mathcal T_\sigma^{(0,2)}(y_{L,i},y_{R,i})P_R+
\mathcal T_\sigma^{(0,2)}(y_{R,i},y_{L,i})P_L
\right] \ ,
\label{UFFS4.02}
\end{align}
where we have set
\begin{align}
F^{(0,2)} &=
10v^2(11e^2-3K^2)
+3(s_{11}+s_{22}+4s_{12}),\\
V^{(0,2)} &=
(F^{(0,2)}+12s_{12})
\left(\bar{\slashed p}_1+\bar{\slashed p}_2\right)
+12(s_{11}-s_{22})
\left(\bar{\slashed p}_1-\bar{\slashed p}_2\right) \, , \\
Q^{(0,2)}_{jk}(x,y) &=
v^2\big[500e^2y+60K^2(x-2y)\big]
\nonumber\\
&+6(x+8y)s_{jj}-6(3x+y)s_{jk}
+6(2x+7y)s_{kk} \, , \\
\mathcal T_\sigma^{(0,2)}(x,y)={}&
xy\big[
Q^{(0,2)}_{12}(x,y)\slashed{\hat p}_1
+Q^{(0,2)}_{21}(y,x)\slashed{\hat p}_2
\big]
\nonumber\\
&+12Kv\,xy(x-y)
\big(
\widehat{s}_{12}
+5\slashed{\hat p}_1\slashed{\hat p}_2
\big)
\nonumber\\
&-6x^2
\big(\slashed{\bar p}_1+\slashed{\bar p}_2\big)
\big[
3(x+2y)\widehat{s}_{12}
+5x\slashed{\hat p}_1\slashed{\hat p}_2
\big]
\nonumber\\
&-3Kv\,x^2\Big[
(19x+y)
\big(
\slashed{\bar p}_1\slashed{\hat p}_1
+\slashed{\bar p}_2\slashed{\hat p}_2
\big)
\nonumber\\
&\hspace{24mm}
+(27y-7x)
\big(
\slashed{\bar p}_1\slashed{\hat p}_2
+\slashed{\bar p}_2\slashed{\hat p}_1
\big)
\Big]
\nonumber\\
&+12xy\slashed{\bar p}_1\slashed{\bar p}_2
\big[
(3x+7y)\slashed{\hat p}_1
+(7x+3y)\slashed{\hat p}_2
\big].
\end{align}
\item $(1,0)$-sector
\begin{align}
\Delta^{(1;1,0)}_{0;\omega(q) \bar \psi_i(p_1) \psi_j(p_2) \sigma(p_3)} & = \frac{i e K^2}{48\pi^2v}
(y_{L,i}-y_{R,i})
\left(Kv+2\slashed{\bar p}_1+2\slashed{\bar p}_2\right)
(P_R-P_L) \, , \\
\Delta^{(1;1,0)}_{1;\omega(q) \bar \psi_i(p_1) \psi_j(p_2) \sigma(p_3)} & = 
    \frac{i e K^2}{6\pi^2v}
\left[
\left(y_{L,i}\slashed{\hat p}_1
+y_{R,i}\slashed{\hat p}_2\right)P_R
+
\left(y_{R,i}\slashed{\hat p}_1
+y_{L,i}\slashed{\hat p}_2\right)P_L
\right].
\label{UFFS4.10}
\end{align}
\item $(0,1)$-sector
\begin{align}
    \Delta^{(1;0,1)}_{0;\omega(q) \bar \psi_i(p_1) \psi_j(p_2) \sigma(p_3)} & = \frac{ie(y_{L,i}-y_{R,i})}{480\pi^2v^3}
\left[
-V^{(0,1)}\left(y_{L,i}^{2}P_R-y_{R,i}^{2}P_L\right) \right .\nonumber \\
& \left .
+Kv\,y_{L,i}y_{R,i}G^{(0,1)} (P_R-P_L)
\right] \, , \\
    \Delta^{(1;0,1)}_{1;\omega(q) \bar \psi_i(p_1) \psi_j(p_2) \sigma(p_3)} & =  \frac{ie}{480\pi^2v^3}
\left[
\mathcal T_\sigma^{(0,1)}(y_{L,i},y_{R,i})P_R
+\mathcal T_\sigma^{(0,1)} (y_{R,i},y_{L,i})P_L
\right]\, ,
\label{UFFS4.01}
\end{align}
with
\begin{align}
F^{(0,1)}={}&50v^2(e^2-K^2)
+3(s_{11}+s_{22})+24s_{12},\\
G^{(0,1)}={}&20v^2(5e^2-12K^2)
+17(s_{11}+s_{22})+63s_{12},\\
V^{(0,1)}={}&F^{(0,1)} \left(\slashed{\bar p}_1+\slashed{\bar p}_2\right)
\nonumber\\
&+12(s_{11}-s_{22})
\left(\slashed{\bar p}_1-\slashed{\bar p}_2\right)\, , \nonumber \\
Q^{(0,1)}_{jk}(x,y)={}&
40v^2\big[13e^2y+4K^2(x-2y)\big]
\nonumber\\
&+12(x+8y)s_{jj}-12(3x+y)s_{jk}
+12(2x+7y)s_{kk} \, , \nonumber \\
\mathcal T_\sigma^{(0,1)}(x,y)&={}
xy\big[
Q^{(0,1)}_{12}(x,y)\slashed{\hat p}_1
+Q^{(0,1)}_{21}(y,x)\slashed{\hat p}_2
\big]
\nonumber\\
&+2Kv\,xy(x-y)
\big(
23\widehat{s}_{12}
+70\slashed{\hat p}_1\slashed{\hat p}_2
\big)
\nonumber\\
&-12x^2
\big(\slashed{\bar p}_1+\slashed{\bar p}_2\big)
\big[
3(x+2y)\widehat{s}_{12}
+5x\slashed{\hat p}_1\slashed{\hat p}_2
\big]
\nonumber\\
&-Kv\,x^2\Big[
(134x+16y)
\big(
\slashed{\bar p}_1\slashed{\hat p}_1
+\slashed{\bar p}_2\slashed{\hat p}_2
\big)
\nonumber\\
&\hspace{24mm}
+(197y-47x)
\big(
\slashed{\bar p}_1\slashed{\hat p}_2
+\slashed{\bar p}_2\slashed{\hat p}_1
\big)
\Big]
\nonumber\\
&+24xy\slashed{\bar p}_1\slashed{\bar p}_2
\big[
(3x+7y)\slashed{\hat p}_1
+(7x+3y)\slashed{\hat p}_2
\big].
\end{align}
\item $(0,0)$-sector
\begin{align}
    \Delta^{(1;0,0)}_{0;\omega(q) \bar \psi_i(p_1) \psi_j(p_2) \sigma(p_3)} & = \frac{ie(y_{L,i}-y_{R,i})}{960\pi^2v^3}
\Big[
V^{(0,0)} \left(y_{L,i}^{2}P_R-y_{R,i}^{2}P_L\right) \nonumber \\
& 
+2Kv\left(30K^2v^2+y_{L,i}y_{R,i}G^{(0,0)} \right)
(P_R-P_L)
\Big]\, , \nonumber \\
\Delta^{(1;0,0)}_{1;\omega(q) \bar \psi_i(p_1) \psi_j(p_2) \sigma(p_3)} & =
-\frac{ie}{480\pi^2v^3}
\left[
\mathcal T_\sigma^{(0,0)} (y_{L,i},y_{R,i})P_R
+\mathcal T_\sigma^{(0,0)} (y_{R,i},y_{L,i})P_L
\right] \, ,
\label{UFFS4.00}
\end{align}
where we have set
\begin{align}
F^{(0,0)} ={}&3(s_{11}+s_{22})+24s_{12}
      -10v^2(e^2+7K^2),\\
G^{(0,0)} ={}&20v^2(e^2+9K^2)
      -11(s_{11}+s_{22})-39s_{12},\\
V^{(0,0)} ={}&F^{(0,0)}\left(\slashed{\bar p}_1+\slashed{\bar p}_2\right)
\nonumber\\
&+12(s_{11}-s_{22})
\left(\slashed{\bar p}_1-\slashed{\bar p}_2\right)
\end{align}
and
\begin{align}
Q^{(0,0)}_{jk}(x,y)={}&
20v^2\big[e^2y+5K^2(x-2y)\big]
\nonumber\\
&+6(x+8y)s_{jj}-6(3x+y)s_{jk}
+6(2x+7y)s_{kk}\, , \nonumber \\
\mathcal T_\sigma^{(0,0)}(x,y)={}&
xy\big[
Q^{(0,0)}_{12}(x,y)\slashed{\hat p}_1
+Q^{(0,0)}_{21}(y,x)\slashed{\hat p}_2
\big]
\nonumber\\
&+2Kv\,xy(x-y)
\big(
17\widehat{s}_{12}
+40\slashed{\hat p}_1\slashed{\hat p}_2
\big)
\nonumber\\
&-6x^2
\big(\slashed{\bar p}_1+\slashed{\bar p}_2\big)
\big[
3(x+2y)\widehat{s}_{12}
+5x\slashed{\hat p}_1\slashed{\hat p}_2
\big]
\nonumber\\
&-Kv\,x^2\Big[
(77x+13y)
\big(
\slashed{\bar p}_1\slashed{\hat p}_1
+\slashed{\bar p}_2\slashed{\hat p}_2
\big)
\nonumber\\
&\hspace{24mm}
+(116y-26x)
\big(
\slashed{\bar p}_1\slashed{\hat p}_2
+\slashed{\bar p}_2\slashed{\hat p}_1
\big)
\Big]
\nonumber\\
&+12xy\slashed{\bar p}_1\slashed{\bar p}_2
\big[
(3x+7y)\slashed{\hat p}_1
+(7x+3y)\slashed{\hat p}_2
\big].
\end{align}
\end{itemize}

We notice the following identities
\begin{align}
& \mathcal T_\sigma^{(0,1)} = \mathcal T_\sigma^{(0,0)} + \mathcal T_\sigma^{(0,2)} \, , \qquad
 Q^{(0,1)} = Q^{(0,0)} + Q^{(0,2)} \, , \qquad 2 V^{(0,1)} = V^{(0,0)} + V^{(0,2)} \, . 
 \label{UFFS4.relations}
\end{align}

\subsubsection{$\Delta^{(1)}_{\omega(q) \bar \psi_i(p_1) \psi_j(p_2) \chi(p_3)}$}

The relevant contributions are
\begin{itemize}
\item $(1,1)$-sector
\begin{align}
    \Delta^{(1;1,1)}_{0;\omega(q) \bar \psi_i(p_1) \psi_j(p_2) \chi(p_3)} & = 0 \, , \\
    \Delta^{(1;1,1)}_{1;\omega(q) \bar \psi_i(p_1) \psi_j(p_2) \chi(p_3)} & = 0 \, .
    \label{UFFS2.11}
\end{align}
\item $(1,0)$-sector
\begin{align}
    \Delta^{(1;1,0)}_{0;\omega(q) \bar \psi_i(p_1) \psi_j(p_2) \chi(p_3)} & = -\frac{7eK^3}{48\pi^2}
(y_{L,i}-y_{R,i})\, (P_L + P_R) \, , \\
    \Delta^{(1;1,0)}_{1;\omega(q) \bar \psi_i(p_1) \psi_j(p_2) \chi(p_3)} & = \frac{eK^2}{6\pi^2v}
\left[
-\left(y_{L,i}\slashed{\hat p}_1
+y_{R,i}\slashed{\hat p}_2\right)P_R
+\left(y_{R,i}\slashed{\hat p}_1
+y_{L,i}\slashed{\hat p}_2\right)P_L
\right] \, .
    \label{UFFS2.10}
\end{align}
\item $(0,1)$-sector
\begin{align}
    \Delta^{(1;0,1)}_{0;\omega(q) \bar \psi_i(p_1) \psi_j(p_2) \chi(p_3)} & = \frac{eK(y_{L,i}-y_{R,i})}{192\pi^2v^2}
\Big[
y_{L,i}y_{R,i}
\left(Q^{(0,1)}
-18\slashed{\bar p}_1\slashed{\bar p}_2\right)
(P_R+P_L)
\\
&+26Kv\left(\slashed{\bar p}_2-\slashed{\bar p}_1\right)
\left(y_{L,i}^{2}P_R+y_{R,i}^{2}P_L\right)
\Big] \, , \\
    \Delta^{(1;0,1)}_{1;\omega(q) \bar \psi_i(p_1) \psi_j(p_2) \chi(p_3)} & = \frac{eK}{96\pi^2v^2}
\left[
\mathcal T_{\chi}^{(0,1)}(y_{L,i},y_{R,i})P_R
-\mathcal T_{\chi}^{(0,1)}(y_{R,i},y_{L,i})P_L
\right] \, ,
    \label{UFFS2.01}
\end{align}
where we have set
\begin{align}
    Q^{(0,1)}
& =8v^2(5e^2+6K^2)-7(s_{11}+s_{22})+2s_{12} \, , \nonumber \\
\mathcal T_{\chi}^{(0,1)}(x,y)={}&
4Kv\,xy(x+y)
\left(\slashed{\hat p}_1+\slashed{\hat p}_2\right)
\nonumber\\
&-x^2(9x-4y)
\left(\slashed{\bar p}_1-\slashed{\bar p}_2\right)
\left(\slashed{\hat p}_1+\slashed{\hat p}_2\right)
\nonumber\\
&-4xy\Big[
2\left(x\widehat{s}_{11}-y\widehat{s}_{22}\right)
+(x-y)\left(
2\widehat{s}_{12}
+\slashed{\hat p}_1\slashed{\hat p}_2
\right)
\Big].
\end{align}
\item $(0,0)$-sector
\begin{align}
    \Delta^{(1;0,0)}_{0;\omega(q) \bar \psi_i(p_1) \psi_j(p_2) \chi(p_3)} & = 
    \nonumber \\
     {}\frac{eK(y_{L,i}-y_{R,i})}{192\pi^2v^2}
\Bigg\{ &
\Big[
28K^2v^2
+y_{L,i}y_{R,i}
\left(
Q^{(0,0)}
+18\slashed{\bar p}_1\slashed{\bar p}_2
\right)
\Big](P_R+P_L)
\\
&\quad
+26Kv
\left(\slashed{\bar p}_1-\slashed{\bar p}_2\right)
\left(y_{L,i}^{2}P_R+y_{R,i}^{2}P_L\right)
\Bigg\} , \\
    \Delta^{(1;0,0)}_{1;\omega(q) \bar \psi_i(p_1) \psi_j(p_2) \chi(p_3)} & = \frac{eK}{96\pi^2v^2}
\left[
\mathcal T_\chi^{(0,0)}(y_{L,i},y_{R,i})P_R
-\mathcal T_\chi^{(0,0)}(y_{R,i},y_{L,i})P_L
\right] \, ,
    \label{UFFS2.00}
\end{align}
where we have set
\begin{align}
Q^{(0,0)}
&=7(s_{11}+s_{22})-2s_{12}-48K^2v^2 \, , \nonumber \\
\mathcal T_{\chi}^{(0,0)}(x,y)&={}
16Kv\left(x\slashed{\hat p}_1+y\slashed{\hat p}_2\right)
\nonumber\\
&-4Kv\,xy(x+y)
\left(\slashed{\hat p}_1+\slashed{\hat p}_2\right)
\nonumber\\
&+x^2(9x-4y)
\left(\slashed{\bar p}_1-\slashed{\bar p}_2\right)
\left(\slashed{\hat p}_1+\slashed{\hat p}_2\right)
\nonumber\\
&+4xy\Big[
2\left(x\widehat{s}_{11}-y\widehat{s}_{22}\right)
+(x-y)\left(
2\widehat{s}_{12}
+\slashed{\hat p}_1\slashed{\hat p}_2
\right)
\Big].
\end{align}
We notice the relation
\begin{align}
    \mathcal T_\chi^{(0,0)}(x,y)
=
-\mathcal T_\chi^{(0,1)}(x,y)
+16Kv\left(
x\slashed{\hat p}_1+y\slashed{\hat p}_2
\right).
\label{UFFS2.relations}
\end{align}
\end{itemize}

\subsection{Ghost-fermion-fermion-vector}

\subsubsection{$\Delta^{(1)}_{\omega(q) \bar \psi_i(p_1) \psi_j(p_2) A_\mu(p_3)}$}

The relevant contributions are
\begin{itemize}
\item $(1,1)$-sector
\begin{align}
    \Delta^{(1;1,1)}_{0;\omega(q) \bar \psi_i(p_1) \psi_j(p_2) A_\mu(p_3)} & = -\frac{ie^2K(y_{L,i}-y_{R,i})}{32\pi^2v}
\Big[
(y_{L,i}-y_{R,i})
(\bar p_1^\mu+\bar p_2^\mu)(P_R+P_L)
    \label{UFFV2.11head}
\\
&\quad+\frac{y_{L,i}+y_{R,i}}{2}
\left[
\bar\gamma^\mu,
\slashed{\bar p}_1+\slashed{\bar p}_2
\right](P_L-P_R)
\Big] \, , \\
    \Delta^{(1;1,1)}_{1;\omega(q) \bar \psi_i(p_1) \psi_j(p_2) A_\mu(p_3)} & = \frac{ie^2K}{8\pi^2v}
(y_{L,i}-y_{R,i})\,
\bar\gamma^\mu
\left(\slashed{\hat p}_1-\slashed{\hat p}_2\right)
\left(y_{L,i}P_R-y_{R,i}P_L\right) \, .
    \label{UFFV2.11}
\end{align}
\item $(1,0)$-sector
\begin{align}
    \Delta^{(1;1,0)}_{0;\omega(q) \bar \psi_i(p_1) \psi_j(p_2) A_\mu(p_3)} & = {}\frac{ie^2K(y_{L,i}-y_{R,i})}{32\pi^2v}
\Big[
(y_{L,i}-y_{R,i})
(\bar p_1^\mu+\bar p_2^\mu)(P_R+P_L)
\\
&\quad+\frac{y_{L,i}+y_{R,i}}{2}
\left[
\bar\gamma^\mu,
\slashed{\bar p}_1+\slashed{\bar p}_2
\right](P_L-P_R)
\Big] \, , \\
    \Delta^{(1;1,0)}_{1;\omega(q) \bar \psi_i(p_1) \psi_j(p_2) A_\mu(p_3)} & = -\frac{ie^2K}{8\pi^2v}
(y_{L,i}-y_{R,i})\,
\bar\gamma^\mu
\left(\slashed{\hat p}_1-\slashed{\hat p}_2\right)
\left(y_{L,i}P_R-y_{R,i}P_L\right) \, .
    \label{UFFV2.10}
\end{align}
\item $(0,1)$-sector
\begin{align}
    \Delta^{(1;0,1)}_{0;\omega(q) \bar \psi_i(p_1) \psi_j(p_2) A_\mu(p_3)} & = \frac{ie^2(y_{L,i}-y_{R,i})}{384\pi^2v^2}
\left[
\mathcal T_0^{\mu}(y_{L,i},y_{R,i})P_R
-\mathcal T_0^{\mu}(y_{R,i},y_{L,i})P_L
\right] \, , 
    \label{UFFV2.01head}\\
    \Delta^{(1;0,1)}_{1;\omega(q) \bar \psi_i(p_1) \psi_j(p_2) A_\mu(p_3)} & = \frac{ie^2}{96\pi^2v^2}
\left[
\mathcal T_1^{\mu}(y_{L,i},y_{R,i})P_R
+\mathcal T_1^{\mu}(y_{R,i},y_{L,i})P_L
\right] \, ,
    \label{UFFV2.01}
\end{align}
where we have set
\begin{align}
   \mathcal T_0^{\mu}(x,y)
={}&x^2\Big[
(5x+y)(s_{22}-s_{11})\bar\gamma^\mu
\nonumber\\
&\quad
-2(4x-3y)
\left(
\bar p_1^\mu\slashed{\bar p}_1
-\bar p_2^\mu\slashed{\bar p}_2
\right)
\nonumber\\
&\quad
+2(9x-4y)
\left(
\bar p_1^\mu\slashed{\bar p}_2
-\bar p_2^\mu\slashed{\bar p}_1
\right)
\Big]
\nonumber\\
&+Kv\,xy\Big[
9(x-y)
\left[
\bar\gamma^\mu,
\slashed{\bar p}_1-\slashed{\bar p}_2
\right]
\nonumber\\
&\hspace{25mm}
-2(x+y)(\bar p_1^\mu-\bar p_2^\mu)
\Big] \, , \nonumber \\
\mathcal T_1^{\mu}(x,y)
={}&x^2(6x-y)\bar\gamma^\mu
\Big[
y(\widehat{s}_{11}-\widehat{s}_{22})
-Kv(x-y)
\left(\slashed{\hat p}_1-\slashed{\hat p}_2\right)
\Big]
\nonumber\\
&+3xy\bar\gamma^\mu\Big[
(x+y)^2
\left(
\slashed{\bar p}_1\slashed{\hat p}_1
+\slashed{\bar p}_2\slashed{\hat p}_2
\right)
\nonumber\\
&\hspace{28mm}
+(2x^2+xy+y^2)
\slashed{\bar p}_1\slashed{\hat p}_2
+(x^2+xy+2y^2)
\slashed{\bar p}_2\slashed{\hat p}_1
\Big]
\nonumber\\
&+xy\Big[
(5y^2-8x^2-8xy)\bar p_1^\mu
+(y^2-8x^2-6xy)\bar p_2^\mu
\Big]\slashed{\hat p}_1
\nonumber\\
&+xy\Big[
(2y^2-13x^2)\bar p_1^\mu
+(2y^2-11x^2-4xy)\bar p_2^\mu
\Big]\slashed{\hat p}_2 \, .
\end{align}
\item $(0,0)$-sector
\begin{align}
    \Delta^{(1;0,0)}_{0;\omega(q) \bar \psi_i(p_1) \psi_j(p_2) A_\mu(p_3)} & = -\frac{ie^2(y_{L,i}-y_{R,i})}{384\pi^2v^2}
\left[
\mathcal T_0^{\mu}(y_{L,i},y_{R,i})P_R
-\mathcal T_0^{\mu}(y_{R,i},y_{L,i})P_L
\right]  \, , 
    \label{UFFV2.00head}
\\
    \Delta^{(1;0,0)}_{1;\omega(q) \bar \psi_i(p_1) \psi_j(p_2) A_\mu(p_3)} & = -\frac{ie^2}{96\pi^2v^2}
\left[
\mathcal T_1^{\mu}(y_{L,i},y_{R,i})P_R
+\mathcal T_1^{\mu}(y_{R,i},y_{L,i})P_L
\right] \,  \, .
    \label{UFFV2.00}
\end{align}
\end{itemize}

\section{One-loop symmetry-restoring counter-terms}\label{app:cts}
\subsection{Scalar and vector sector}
Since by integration by parts the ghost is undifferentiated in the ST breaking, in this sector the most general
ST breaking term of dim.$\leq 5$ can be written as
\begin{align}
    {\cal B}= \int d^D x \, \omega P(\sigma,\chi,A_\mu)
\end{align}
where $P(\sigma,\chi,A_\mu)$ is a C-odd polynomial in the fields $\sigma,\chi,A_\mu$ and their derivatives.
In order to keep the notation compact we introduce an index $a$ running over the subspace $B$ of barred covariants and the orthogonal subspace $H$ of hatted (evanescent) covariants.
I.e. we set
\begin{align}
\partial_{\mathrm B}&=\bar\partial,&
\partial_{\mathrm H}&=\hat\partial,&
A_{\mathrm B}&=\bar A,& A_{\mathrm H}&=\hat A,\\
\square_a&=\partial_{a\mu}\partial_a^\mu,&
d_a&=\partial_{a\mu}A_a^\mu,&
T_{ab}&=A_a^\mu A_b^\nu\partial_{a\mu}A_{b\nu} \, ,
\end{align}
with $a,b\in\{\mathrm B,\mathrm H\}$.
Then the most general form of $P$ can be written as (a dot denotes the Lorentz contraction in a given subspace, e.g.
$A_a\cdot\partial_a\sigma$ equals
$\bar A_\mu \bar \partial^\mu \sigma$ for $a=B$,
$\widehat{A}_\mu \widehat{\partial}^\mu \sigma$ for $a=H$)
\begin{align}
P={}&V_\Delta
 +\sum_a\bigl[
 \ell_a\square_a\chi+b_a\chi\square_a\sigma
 +c_a\sigma\square_a\chi
 +e_a\partial_a\chi\cdot\partial_a\sigma\bigr]\nonumber\\
&+\sum_a\bigl[
 (u_a+p_a\sigma+r_a\sigma^2+z_a\chi^2)d_a
 +(w_a+t_a\sigma)A_a\cdot\partial_a\sigma
 +h_a\chi A_a\cdot\partial_a\chi\bigr]\nonumber\\
&+\sum_a(m_a\chi+n_a\chi\sigma)A_a^2
 +\sum_{a,b}\bigl[k_{ab}\square_a d_b
                 +j_{ab}d_a A_b^2+\kappa_{ab}T_{ab}\bigr],
\label{eq:b-PD}\\
V_\Delta={}&\alpha_1\chi+\alpha_2\chi\sigma
 +\alpha_3\chi\sigma^2+\alpha_4\chi^3
 +\alpha_5\chi\sigma^3+\alpha_6\chi^3\sigma.
\label{eq:b-Vdelta}
\end{align}
The coefficients in the above equation are not all independent. In fact the Wess-Zumino consistency condition
\begin{align}
s {\cal B} = 0
\end{align}
entails the following relations for $a\in \{B, H\}$:
\begin{align}
w_a&=p_a-e\ell_a-ev(e_a-c_a),\label{eq:b-condw}\\
t_a&=2r_a-ee_a,\label{eq:b-condt}\\
h_a&=2z_a+ee_a,\label{eq:b-condh}\\
m_a&=e\left(\frac{p_a}{2}-\frac{e\ell_a}{2}
        -vz_a-eve_a+\frac{evc_a}{2}\right),\label{eq:b-condm}\\
n_a&=e(r_a-z_a-e e_a).\label{eq:b-condn}
\end{align}
Moreover 
\begin{equation}\label{eq:b-condj}
j_{ab}=j_{ba},\qquad \kappa_{ab}=2j_{ab}.
\end{equation}
Under these conditions we can parameterize a BRST inverse $\Xi$ for ${\cal B}$ as follows.
Let us define the polynomial
\begin{align}
V_\Xi={}&\frac{\alpha_1}{e}\sigma
 +\frac{\alpha_2+2v\alpha_4}{2e}\sigma^2
 +\frac{\alpha_3+2\alpha_4+v\alpha_6}{3e}\sigma^3\nonumber\\
&+\frac{\alpha_5+\alpha_6}{4e}\sigma^4
 +\frac{\alpha_4}{e}\sigma\chi^2
 +\frac{\alpha_6}{2e}\sigma^2\chi^2,
\label{eq:b-Vxi}
\end{align}
and for each $a \in \{B,H \}$ let us define the polynomials
\begin{align}
f_a(\sigma,\chi)&=f_{0a}+f_{1a}\sigma+f_{2a}\sigma^2+f_{\chi a}\chi^2,
\label{eq:b-fpoly}\\
f_{0a}&=u_a-ev\ell_a,&f_{1a}&=p_a-e\ell_a-evc_a,\nonumber\\
f_{2a}&=r_a-ec_a,&f_{\chi a}&=z_a+e(e_a-c_a).
\label{eq:b-fcoeff}
\end{align}
Then one particular inverse
is given by
\begin{align}
\Xi=\int d^D x\,\Bigg\{&V_\Xi
 +\sum_a\left[
 \frac{e_a-c_a-b_a}{2e}(\partial_a\sigma)^2
 +\ell_a A_a\cdot\partial_a\chi\right]\nonumber\\
&+\sum_a\left[
 c_a\sigma A_a\cdot\partial_a\chi
 +(e_a-c_a)\chi A_a\cdot\partial_a\sigma
 +\frac12 A_a^2 f_a(\sigma,\chi)\right]\nonumber\\
&-\frac12\sum_{a,b}k_{ab}
  (\partial_{a\mu}A_{b\nu})(\partial_a^\mu A_b^\nu)
 +\frac14\sum_{a,b}j_{ab}A_a^2A_b^2\Bigg\}.
\label{eq:b-XiD}
\end{align}
The condition $s \Xi = - {\cal B}$ can then be checked by direct computation once one imposes Eqs.(\ref{eq:b-condw})-(\ref{eq:b-condj}).

\subsection{Fermionic sector}
In the fermionic sector without derivatives there are two possible breaking terms of dimension $\leq 5$,
namely
\begin{align}
    {\cal B}^{\cal J}_{\omega \bar \psi_i \psi_i} \equiv \int d^Dx \,  \omega \bar \psi_i P_{\cal J}\psi_i \, , \qquad {\cal J}= L,R \, .
\end{align}
One BRST inverse ${\Xi}^{\cal J}_{\omega \bar \psi_i \psi_i}$ fulfilling $ s{\Xi}^{\cal J}_{\omega \bar \psi_i \psi_i} = - {\cal B}^{\cal J}_{\omega \bar \psi_i \psi_i}$ is easily found to be 
\begin{align}
{\Xi}^{\cal J}_{\omega \bar \psi_i \psi_i} = 
\frac{i \epsilon_{\cal J}}{e (y_{L,i}-y_{R,i})} \int d^D x \,
\bar \psi_i P_{\cal J} \psi_i \, , \qquad
\epsilon_L=-\epsilon_R=1 \, .
\end{align}
We now move to the sector with one derivative. For that purpose let us define the following  fermionic bilinears
\begin{align}
H_{{\cal J},i}&=(\bar\partial_\mu\bar\psi_i)\bar\gamma^\mu P_{\cal J}\psi_i,
&K_{{\cal J},i}&=\bar\psi_i\bar\gamma^\mu P_{\cal J}\bar\partial_\mu\psi_i,\\
\widehat H_{{\cal J},i}&=(\hat\partial_\mu\bar\psi_i)\hat\gamma^\mu P_{\cal J}\psi_i,
&\widehat K_{{\cal J},i}&=\bar\psi_i\hat\gamma^\mu P_{\cal J}\hat\partial_\mu\psi_i,\\
J_{{\cal J},i}^\mu&=\bar\psi_i\bar\gamma^\mu P_{\cal J}\psi_i,
&\widehat J_{{\cal J},i}^\mu&=\bar\psi_i\hat\gamma^\mu P_{\cal J}\psi_i \, .
\label{ffder}
\end{align}
Their BRST variations read
\begin{align}
& sH_{{\cal J},i}=-ie y_{{\cal J},i}(\bar\partial_\mu\omega)J_{{\cal J},i}^\mu,\qquad 
 sK_{{\cal J},i}= i e y_{{\cal J},i}(\bar\partial_\mu\omega)J_{{\cal J},i}^\mu,\nonumber \\
& s\widehat H_{{\cal J},i}=i\delta_{{\cal J},i} \omega\widehat H_{{\cal J},i}
 -ie y_{\bar {\cal J},i}(\hat\partial_\mu\omega)\widehat J_{\cal J,i}^\mu,\qquad
s\widehat K_{{\cal J},i}=i\delta_{{\cal J},i}\omega\widehat K_{{\cal J},i}
 +ie y_{{\cal J},i}(\hat\partial_\mu\omega)\widehat J_{{\cal J},i}^\mu \, ,
\label{brst.ffder}
\end{align}
where we have set $\delta_{{\cal J},i} = e (y_{{\cal J},i} -y_{\bar {\cal J},i})$ and $\bar {L} = R, \bar {R} = L$.

The most general breaking in this sector can be parameterized as
\begin{equation}\label{eq:Delta}
{\cal B}_{\omega \partial \bar \psi_i \psi_i}=\int d^D x\,\omega\sum_{{\cal J}=L,R}
 \left(a_{{\cal J},i} H_{{\cal J},i}+b_{{\cal J},i}K_{{\cal J},i}+\widehat a_{{\cal J},i}\widehat H_{{\cal J},i}
                         +\widehat b_{{\cal J},i}\widehat K_{{\cal J},i}\right).
\end{equation}
The coefficients $a,b,\widehat{a},\widehat{b}$'s are not independent and must obey the linear relations imposed by
the Wess-Zumino consistency condition $s {\cal B}_{\omega \partial \bar \psi_i \psi_i} = 0$. 

By using Eq.(\ref{brst.ffder})
one finds the following conditions (we assume $y_{L,i}y_{R,i}\neq 0$):
\begin{equation}\label{eq:closure-nonzero}
a_{L,i}=b_{L,i},\qquad a_{R,i}=b_{R,i},\qquad
y_{L,i}\widehat b_{L,i}=y_{R,i}\widehat a_{L,i},\qquad
y_{R,i}\widehat b_{R,i}=y_{L,i} \widehat a_{R,i}.
\end{equation}
Then by direct computation one gets the following
BRST-inverse
\begin{align}
\label{eq:Xi-kinetic}
\Xi_{\omega \partial \bar \psi_i \psi_i}=-i\int d^D x\sum_{X=L,R}
 \left(\frac{a_{X,i}}{e y_{X,i} }K_{X,i} +
       \frac{\widehat a_{X,i}}{e y_{X,i}}\widehat K_{X,i} \right).
\end{align}
Finally in the sector with one ghost, two fermions and one scalar the relevant
most general ST breaking of dimension $\leq 5$ can be parameterized as
\begin{align}
    {\cal B}_{\omega \bar \psi_i \psi_i \Phi} = 
    \int d^D x \, \omega \Big ( d_{L,i} (\sigma - i \chi) \bar \psi_i P_L \psi_i + 
    d_{R,i} (\sigma + i \chi) \bar \psi_i P_R \psi_i \Big ) \, .
\end{align}
Then an explicit inverse is
\begin{align}
    {\Xi}_{\omega \bar \psi_i \psi_i \Phi} = 
    \frac{i}{2e} \int d^D x \, \Big [ - d_{L,i} ( \sigma - v - i \chi) \bar \psi_i P_L \psi_i +
    d_{R,i} ( \sigma - v + i \chi) \bar \psi_i P_R \psi_i  \Big ] \, ,
\end{align}
as can be seen directly by noticing the relations
\begin{align}
s ( ( \sigma - v - i \epsilon_{\cal J} \chi) \bar \psi_i P_{\cal J} \psi_i ) = -2i \epsilon_{\cal J} e \omega (\sigma - i \epsilon_{\cal J} \chi) \bar \psi_i P_{\cal J}  \psi_i \, , \qquad {\cal J}=L,R \, .
\end{align}
\bibliography{bibliography_1loop}

@article{Binosi:2017ubk,
    author = "Binosi, Daniele and Quadri, Andrea",
    title = "{Off-shell renormalization in Higgs effective field theories}",
    eprint = "1709.09937",
    archivePrefix = "arXiv",
    primaryClass = "hep-th",
    doi = "10.1007/JHEP04(2018)050",
    journal = "JHEP",
    volume = "04",
    pages = "050",
    year = "2018"
}

@article{Binosi:2020unh,
    author = "Binosi, D. and Quadri, A.",
    title = "{Off-shell renormalization in the presence of dimension 6 derivative operators. Part III. Operator mixing and \ensuremath{\beta} functions}",
    eprint = "2001.07430",
    archivePrefix = "arXiv",
    primaryClass = "hep-ph",
    doi = "10.1007/JHEP05(2020)141",
    journal = "JHEP",
    volume = "05",
    pages = "141",
    year = "2020"
}

@article{Barnich:2000zw,
      author         = "Barnich, Glenn and Brandt, Friedemann and Henneaux, Marc",
      title          = "{Local BRST cohomology in gauge theories}",
      journal        = "Phys. Rept.",
      volume         = "338",
      year           = "2000",
      pages          = "439-569",
      doi            = "10.1016/S0370-1573(00)00049-1",
      eprint         = "hep-th/0002245",
      archivePrefix  = "arXiv",
      primaryClass   = "hep-th",
      reportNumber   = "ITP-UH-03-00, ULB-TH-00-05",
      SLACcitation   = "%%CITATION = HEP-TH/0002245;%%"
}

@article{Hahn:2000kx,
      author         = "Hahn, Thomas",
      title          = "{Generating Feynman diagrams and amplitudes with FeynArts
                        3}",
      journal        = "Comput. Phys. Commun.",
      volume         = "140",
      year           = "2001",
      pages          = "418-431",
      doi            = "10.1016/S0010-4655(01)00290-9",
      eprint         = "hep-ph/0012260",
      archivePrefix  = "arXiv",
      primaryClass   = "hep-ph",
      reportNumber   = "KA-TP-23-2000",
      SLACcitation   = "%%CITATION = HEP-PH/0012260;%%"
}

@article{Gomis:1994he,
      author         = "Gomis, Joaquim and Paris, Jordi and Samuel, Stuart",
      title          = "{Antibracket, antifields and gauge theory quantization}",
      journal        = "Phys. Rept.",
      volume         = "259",
      year           = "1995",
      pages          = "1-145",
      doi            = "10.1016/0370-1573(94)00112-G",
      eprint         = "hep-th/9412228",
      archivePrefix  = "arXiv",
      primaryClass   = "hep-th",
      reportNumber   = "CCNY-HEP-94-03, KUL-TF-94-12, UB-ECM-PF-94-15,
                        UTTG-11-94",
      SLACcitation   = "%%CITATION = HEP-TH/9412228;%%"
}

@article{Ferrari:1999nj,
      author         = "Ferrari, Ruggero and Grassi, Pietro Antonio and Quadri,
                        Andrea",
      title          = "{Direct algebraic restoration of Slavnov-Taylor
                        identities in the Abelian Higgs-Kibble model}",
      journal        = "Phys. Lett.",
      volume         = "B472",
      year           = "2000",
      pages          = "346-356",
      doi            = "10.1016/S0370-2693(99)01452-5",
      eprint         = "hep-th/9905192",
      archivePrefix  = "arXiv",
      primaryClass   = "hep-th",
      reportNumber   = "IFUM-641-FT, MPI-PHT-99-20",
      SLACcitation   = "%%CITATION = HEP-TH/9905192;%%"
}

@article{Piguet:1995er,
      author         = "Piguet, O. and Sorella, S. P.",
      title          = "{Algebraic renormalization: Perturbative renormalization,
                        symmetries and anomalies}",
      journal        = "Lect. Notes Phys. Monogr.",
      volume         = "28",
      year           = "1995",
      pages          = "1-134",
      doi            = "10.1007/978-3-540-49192-7",
      SLACcitation   = "%%CITATION = INSPIRE-405127;%%"
}

@article{Quadri:2005pv,
      author         = "Quadri, Andrea",
      title          = "{Slavnov-Taylor parameterization of Yang-Mills theory
                        with massive fermions in the presence of singlet
                        axial-vector currents}",
      journal        = "JHEP",
      volume         = "06",
      year           = "2005",
      pages          = "068",
      doi            = "10.1088/1126-6708/2005/06/068",
      eprint         = "hep-th/0504076",
      archivePrefix  = "arXiv",
      primaryClass   = "hep-th",
      reportNumber   = "IFUM-824-FT",
      SLACcitation   = "%%CITATION = HEP-TH/0504076;%%"
}

@article{Breitenlohner:1977hr,
      author         = "Breitenlohner, P. and Maison, D.",
      title          = "{Dimensional Renormalization and the Action Principle}",
      journal        = "Commun. Math. Phys.",
      volume         = "52",
      year           = "1977",
      pages          = "11-38",
      doi            = "10.1007/BF01609069",
      SLACcitation   = "%%CITATION = CMPHA,52,11;%%"
}

@article{Ferrari:2000yp,
      author         = "Ferrari, Ruggero and Picariello, Marco and Quadri,
                        Andrea",
      title          = "{Algebraic aspects of the background field method}",
      journal        = "Annals Phys.",
      volume         = "294",
      year           = "2001",
      pages          = "165-181",
      doi            = "10.1006/aphy.2001.6198",
      eprint         = "hep-th/0012090",
      archivePrefix  = "arXiv",
      primaryClass   = "hep-th",
      reportNumber   = "IFUM-669-FT",
      SLACcitation   = "%%CITATION = HEP-TH/0012090;%%"
}

@article{Grassi:1995wr,
      author         = "Grassi, P. A.",
      title          = "{Stability and renormalization of Yang-Mills theory with
                        background field method: A Regularization independent
                        proof}",
      journal        = "Nucl. Phys.",
      volume         = "B462",
      year           = "1996",
      pages          = "524-550",
      doi            = "10.1016/0550-3213(96)00017-X",
      eprint         = "hep-th/9505101",
      archivePrefix  = "arXiv",
      primaryClass   = "hep-th",
      reportNumber   = "IFUM-452-FTA",
      SLACcitation   = "%%CITATION = HEP-TH/9505101;%%"
}

@article{Becchi:1999ir,
      author         = "Becchi, Carlo and Collina, Renzo",
      title          = "{Further comments on the background field method and
                        gauge invariant effective actions}",
      journal        = "Nucl. Phys.",
      volume         = "B562",
      year           = "1999",
      pages          = "412-430",
      doi            = "10.1016/S0550-3213(99)00555-6",
      eprint         = "hep-th/9907092",
      archivePrefix  = "arXiv",
      primaryClass   = "hep-th",
      reportNumber   = "GEF-TH-04-994",
      SLACcitation   = "%%CITATION = HEP-TH/9907092;%%"
}

@article{Quadri:2003pq,
      author         = "Quadri, Andrea",
      title          = "{Higher order nonsymmetric counterterms in pure
                        Yang-Mills theory}",
      journal        = "J. Phys.",
      volume         = "G30",
      year           = "2004",
      pages          = "677",
      doi            = "10.1088/0954-3899/30/5/009",
      eprint         = "hep-th/0309133",
      archivePrefix  = "arXiv",
      primaryClass   = "hep-th",
      reportNumber   = "MPI-PHT-2003-22, MPP-2003-71",
      SLACcitation   = "%%CITATION = HEP-TH/0309133;%%"
}

@article{Grassi:2001zz,
      author         = "Grassi, P. A. and Hurth, T. and Steinhauser, M.",
      title          = "{The Algebraic method}",
      journal        = "Nucl. Phys.",
      volume         = "B610",
      year           = "2001",
      pages          = "215-250",
      doi            = "10.1016/S0550-3213(01)00303-0",
      eprint         = "hep-ph/0102005",
      archivePrefix  = "arXiv",
      primaryClass   = "hep-ph",
      reportNumber   = "NYU-TH-00-09-09, CERN-TH-2001-002, DESY-01-008",
      SLACcitation   = "%%CITATION = HEP-PH/0102005;%%"
}

@article{Belusca-Maito:2023wah,
    author = "B\'elusca-Maito, Hermes and Ilakovac, Amon and Kuehler, Paul and Mador-Bovzinovic, Marija and Stoeckinger, Dominik and Weisswange, Matthias",
    title = "{Introduction to Renormalization Theory and Chiral Gauge Theories in Dimensional Regularization with Non-Anticommuting \ensuremath{\gamma}$_{5}$}",
    eprint = "2303.09120",
    archivePrefix = "arXiv",
    primaryClass = "hep-ph",
    doi = "10.3390/sym15030622",
    journal = "Symmetry",
    volume = "15",
    number = "3",
    pages = "622",
    year = "2023"
}

@article{Cornella:2022hkc,
    author = "Cornella, Claudia and Feruglio, Ferruccio and Vecchi, Luca",
    title = "{Gauge invariance and finite counterterms in chiral gauge theories}",
    eprint = "2205.10381",
    archivePrefix = "arXiv",
    primaryClass = "hep-ph",
    doi = "10.1007/JHEP02(2023)244",
    journal = "JHEP",
    volume = "02",
    pages = "244",
    year = "2023"
}

@article{Shtabovenko:2020gxv,
    author = "Shtabovenko, Vladyslav and Mertig, Rolf and Orellana, Frederik",
    title = "{FeynCalc 9.3: New features and improvements}",
    eprint = "2001.04407",
    archivePrefix = "arXiv",
    primaryClass = "hep-ph",
    reportNumber = "P3H-20-002, TTP19-020, TUM-EFT 130/19",
    doi = "10.1016/j.cpc.2020.107478",
    journal = "Comput. Phys. Commun.",
    volume = "256",
    pages = "107478",
    year = "2020"
}

@article{Shtabovenko:2016sxi,
    author = "Shtabovenko, Vladyslav and Mertig, Rolf and Orellana, Frederik",
    title = "{New Developments in FeynCalc 9.0}",
    eprint = "1601.01167",
    archivePrefix = "arXiv",
    primaryClass = "hep-ph",
    reportNumber = "TUM-EFT-71-15",
    doi = "10.1016/j.cpc.2016.06.008",
    journal = "Comput. Phys. Commun.",
    volume = "207",
    pages = "432--444",
    year = "2016"
}

@article{Quadri:2024aqo,
    author = "Quadri, Andrea",
    title = "{Gauge-invariant quantum fields}",
    eprint = "2402.01016",
    archivePrefix = "arXiv",
    primaryClass = "hep-th",
    doi = "10.1140/epjc/s10052-024-13317-0",
    journal = "Eur. Phys. J. C",
    volume = "84",
    number = "9",
    pages = "975",
    year = "2024",
    note = "[Erratum: Eur.Phys.J.C 84, 1073 (2024)]"
}

@article{Martin:1999cc,
    author = "Martin, C. P. and Sanchez-Ruiz, D.",
    title = "{Action principles, restoration of BRS symmetry and the renormalization group equation for chiral nonAbelian gauge theories in dimensional renormalization with a nonanticommuting gamma(5)}",
    eprint = "hep-th/9905076",
    archivePrefix = "arXiv",
    reportNumber = "FT-UCM-20-99",
    doi = "10.1016/S0550-3213(99)00453-8",
    journal = "Nucl. Phys. B",
    volume = "572",
    pages = "387--477",
    year = "2000"
}

@article{Sanchez-Ruiz:2002pcf,
    author = "Sanchez-Ruiz, D.",
    title = "{BRS symmetry restoration of chiral Abelian Higgs-Kibble theory in dimensional renormalization with a nonanticommuting gamma(5)}",
    eprint = "hep-th/0209023",
    archivePrefix = "arXiv",
    doi = "10.1103/PhysRevD.68.025009",
    journal = "Phys. Rev. D",
    volume = "68",
    pages = "025009",
    year = "2003"
}

@article{Ebert:2024xpy,
    author = {Ebert, Paul L. and K{\"u}hler, Paul and St{\"o}ckinger, Dominik and Wei{\ss}wange, Matthias},
    title = "{Shedding light on evanescent shadows {\textemdash} Exploration of non-anticommuting {\ensuremath{\gamma}}$_{5}$ in Dimensional Regularisation}",
    eprint = "2411.02543",
    archivePrefix = "arXiv",
    primaryClass = "hep-ph",
    doi = "10.1007/JHEP01(2025)114",
    journal = "JHEP",
    volume = "01",
    pages = "114",
    year = "2025"
}

@article{Stockinger:2023ndm,
    author = {St{\"o}ckinger, Dominik and Wei{\ss}wange, Matthias},
    title = "{Full three-loop renormalisation of an abelian chiral gauge theory with non-anticommuting $\gamma_{5}$ in the BMHV scheme}",
    eprint = "2312.11291",
    archivePrefix = "arXiv",
    primaryClass = "hep-ph",
    doi = "10.1007/JHEP02(2024)139",
    journal = "JHEP",
    volume = "02",
    pages = "139",
    year = "2024"
}

@article{Kuhler:2025znv,
    author = {K{\"u}hler, Paul and St{\"o}ckinger, Dominik},
    title = "{Two-loop renormalization of a chiral SU(2) gauge theory in dimensional regularization with non-anticommuting {\ensuremath{\gamma}}$_{5}$}",
    eprint = "2504.06080",
    archivePrefix = "arXiv",
    primaryClass = "hep-ph",
    doi = "10.1007/JHEP07(2025)082",
    journal = "JHEP",
    volume = "07",
    pages = "082",
    year = "2025"
}

@article{vonManteuffel:2025swv,
    author = {von Manteuffel, Andreas and St{\"o}ckinger, Dominik and Wei{\ss}wange, Matthias},
    title = "{Four-loop renormalisation of chiral gauge theories with non-anticommuting {\ensuremath{\gamma}}$_{5}$ in the BMHV scheme}",
    eprint = "2506.12253",
    archivePrefix = "arXiv",
    primaryClass = "hep-ph",
    doi = "10.1007/JHEP08(2025)088",
    journal = "JHEP",
    volume = "08",
    pages = "088",
    year = "2025"
}

@article{Shtabovenko:2023idz,
    author = "Shtabovenko, Vladyslav and Mertig, Rolf and Orellana, Frederik",
    title = "{FeynCalc 10: Do multiloop integrals dream of computer codes?}",
    eprint = "2312.14089",
    archivePrefix = "arXiv",
    primaryClass = "hep-ph",
    reportNumber = "P3H-23-089, TTP23-056, SI-HEP-2023-27",
    doi = "10.1016/j.cpc.2024.109357",
    journal = "Comput. Phys. Commun.",
    volume = "306",
    pages = "109357",
    year = "2025"
}

@article{Mertig:1990an,
    author = "Mertig, R. and Bohm, M. and Denner, Ansgar",
    title = "{FEYN CALC: Computer algebraic calculation of Feynman amplitudes}",
    reportNumber = "PRINT-90-0639 (WURZBURG)",
    doi = "10.1016/0010-4655(91)90130-D",
    journal = "Comput. Phys. Commun.",
    volume = "64",
    pages = "345--359",
    year = "1991"
}

@misc{OpenAIChatGPT2026,
  author       = {{OpenAI}},
  title        = {{Astra}},
  year         = {2026},
  howpublished = {\url{https://chatgpt.com/}},
  note         = {Accessed: 20 September 2026}
}

@article{Chetyrkin:1997fm,
    author = "Chetyrkin, Konstantin G. and Misiak, Mikolaj and Munz, Manfred",
    title = "{Beta functions and anomalous dimensions up to three loops}",
    eprint = "hep-ph/9711266",
    archivePrefix = "arXiv",
    reportNumber = "MPI-PHT-97-45, TTP-97-43, ZU-TH-16-97, TUM-HEP-284-97, IFT-11-97",
    doi = "10.1016/S0550-3213(98)00122-9",
    journal = "Nucl. Phys. B",
    volume = "518",
    pages = "473--494",
    year = "1998"
}

@article{Lang:2021hnw,
    author = "Lang, Jean-Nicolas and Pozzorini, Stefano and Zhang, Hantian and Zoller, Max F.",
    title = "{Two-loop rational terms for spontaneously broken theories}",
    eprint = "2107.10288",
    archivePrefix = "arXiv",
    primaryClass = "hep-ph",
    reportNumber = "PSI-PR-21-14, ZU-TH 31/21",
    doi = "10.1007/JHEP01(2022)105",
    journal = "JHEP",
    volume = "01",
    pages = "105",
    year = "2022"
}

@article{OlgosoRuiz:2024dzq,
    author = "Olgoso Ruiz, Pablo and Vecchi, Luca",
    title = "{Spurious gauge-invariance and {\ensuremath{\gamma}}$_{5}$ in dimensional regularization}",
    eprint = "2406.17013",
    archivePrefix = "arXiv",
    primaryClass = "hep-ph",
    doi = "10.1007/JHEP12(2024)080",
    journal = "JHEP",
    volume = "12",
    pages = "080",
    year = "2024"
}

@article{Fuentes-Martin:2025meq,
    author = "Fuentes-Mart{\'\i}n, Javier and Moreno-S{\'a}nchez, Adri{\'a}n and Thomsen, Anders Eller",
    title = "{Symmetry restoration in the SMEFT: finite counterterms for a non-anticommuting gamma-5}",
    eprint = "2507.19589",
    archivePrefix = "arXiv",
    primaryClass = "hep-ph",
    doi = "10.1007/JHEP02(2026)150",
    journal = "JHEP",
    volume = "02",
    pages = "150",
    year = "2026"
}

@phdthesis{Weisswange:2025tab,
    author = "Wei{\ss}wange, Matthias",
    title = "{Renormalisation of Chiral Gauge Theories with Non-Anticommuting {\ensuremath{\gamma}}5 at the Multi-Loop Level}",
    eprint = "2603.05059",
    archivePrefix = "arXiv",
    primaryClass = "hep-ph",
    school = {Technische Universit{\"a}t Dresden, Institut f{\"u}r Kern- und Teilchenphysik, Dresden, Tech. U.},
    year = "2025"
}

\end{document}